\documentclass[aps,prm,twocolumn,superscriptaddress,floatfix]{revtex4-1}

\usepackage[latin1]{inputenc}
\usepackage{tabularx}
\usepackage{graphicx}
\usepackage{dcolumn}
\usepackage{bm}
\usepackage{amssymb}
\usepackage{microtype}
\usepackage{xfrac}
\usepackage{makecell}
\usepackage{array}
\usepackage[version=4]{mhchem}
\usepackage{gensymb}
\usepackage{multirow}
\usepackage[colorlinks=true]{hyperref}

\newcommand{\angstrom}{\text{\normalfont\AA}}

\makeatletter
\newcommand\footnoteref[1]{\protected@xdef\@thefnmark{\ref{#1}}\@footnotemark}
\makeatother

\begin{document}
\title{Stripe Helical Magnetism and Two Regimes of Anomalous Hall Effect in \ce{NdAlGe}}

\author{Hung-Yu~Yang}
\thanks{Present address: Department of Electrical and Computer Engineering, University of California, Los Angeles, California 90095, USA}
\thanks{hungyuyang@g.ucla.edu}
\affiliation{Department of Physics, Boston College, Chestnut Hill, MA 02467, USA}

\author{Jonathan~Gaudet}
\affiliation{NIST Center for Neutron Research, National Institute of Standards and Technology, Gaithersburg, Maryland 20899, USA}
\affiliation{Department of Materials Science and Eng., University of Maryland, College Park, MD 20742-2115}

\author{Rahul~Verma}
\affiliation{Department of Condensed Matter Physics and Materials Science, Tata Institute of Fundamental Research, Colaba, Mumbai 400005, India}

\author{Santu~Baidya}
\affiliation{Department of Physics and Astronomy, Rutgers University, Piscataway, New Jersey 08854-8019, USA}
\affiliation{Department of Physics and Materials Science, Jaypee University of Information Technology, Waknaghat, Solan, Himachal Pradesh 173234, India}

\author{Faranak~Bahrami}
\affiliation{Department of Physics, Boston College, Chestnut Hill, MA 02467, USA}

\author{Xiaohan~Yao}
\affiliation{Department of Physics, Boston College, Chestnut Hill, MA 02467, USA}

\author{Cheng-Yi~Huang}
\affiliation{Department of Physics, Northeastern University, Boston, MA 02115, USA}

\author{Lisa~DeBeer-Schmitt}
\affiliation{Neutron Scattering Division, Oak Ridge National Laboratory, Oak Ridge, Tennessee 37831, USA}

\author{Adam~A.~Aczel}
\affiliation{Neutron Scattering Division, Oak Ridge National Laboratory, Oak Ridge, Tennessee 37831, USA}

\author{Guangyong~Xu}
\affiliation{NIST Center for Neutron Research, National Institute of Standards and Technology, Gaithersburg, Maryland 20899, USA}

\author{Hsin~Lin}
\affiliation{Institute of Physics, Academia Sinica, Taipei 115201, Taiwan}

\author{Arun~Bansil}
\affiliation{Department of Physics, Northeastern University, Boston, MA 02115, USA}

\author{Bahadur~Singh}
\affiliation{Department of Condensed Matter Physics and Materials Science, Tata Institute of Fundamental Research, Colaba, Mumbai 400005, India}

\author{Fazel~Tafti}
\affiliation{Department of Physics, Boston College, Chestnut Hill, MA 02467, USA}

\begin{abstract} 
We report the magnetic and electronic transport properties of the inversion and time-reversal symmetry breaking Weyl semimetal NdAlGe. This material is analogous to NdAlSi, whose helical magnetism presents a rare example of a Weyl-mediated collective phenomenon, but with a larger spin-orbit coupling. Our neutron diffraction experiments revealed that NdAlGe, similar to NdAlSi, supports an incommensurate spin density wave ($T_{\text{inc}}=6.8$ K) whose spins are predominantly pointing along the out-of-plane direction and have a small helical spin canting of 3$\degree$. The spin density wave has a long-wavelength of $\sim$ 35~nm and transitions to a commensurate ferrimagnetic state below $T_{\text{com}}=5.1$ K. Using small-angle neutron scattering, we showed that the zero-field cooled ferrimagnetic domains form stripes in real space with characteristic length scales of 18~nm and 72~nm parallel and perpendicular to the [110] direction, respectively. Interestingly, for the transport properties, NdAlSi does not exhibit an anomalous Hall effect (AHE) that is commonly observed in magnetic Weyl semimetals. In contrast to NdAlSi, we identify two different AHE regimes in NdAlGe that are respectively governed by intrinsic Berry curvature and extrinsic disorders/spin fluctuations. Our study suggests that Weyl-mediated magnetism prevails in this group of noncentrosymmetric magnetic Weyl semimetals NdAl$X$, but transport properties including AHE are affected by material-specific extrinsic effects such as disorders, despite the presence of prominent Berry curvature.


\end{abstract}

\date{\today}

\maketitle
\section{Introduction}
To establish a Weyl semimetal phase, one needs to break either inversion or time-reversal symmetry to split Weyl nodes of opposite chirality, which may then lead to interesting topological properties \cite{Wan2011,Armitage2018a}. Both routes have been explored through candidate materials that break either the inversion symmetry, such as the noncentrosymmetric TaAs \cite{Huang2015e,Weng2015,lv_experimental_2015,yang_weyl_2015,Xu2015a}, or time-reversal symmetry, such as the ferromagnetic (FM) Co$_3$Sn$_2$S$_2$ \cite{Liu2018b,Wang2018b}. Weyl semimetals that break \emph{both} inversion and time-reversal symmetries remain largely unexplored, despite theoretical predictions of Weyl-mediated interactions with rich phase diagrams and topological magnetic textures~\cite{Chang2015a,Hosseini2015,Wang2017,Nikolic2020,Nikolic2020a}. For instance, in the $R$Al$X$ ($R=\text{rare-earths}$, $X=\text{Ge/Si}$) material family of double-symmetry-breaking Weyl semimetals~\cite{Chang2018,Xu2017e,Sanchez2020a}, a variety of rich magnetic orders have been found. These include collinear FM order~\cite{Meng2019b,Destraz2020,Sanchez2020a,Yang2020d} and noncollinear FM order~\cite{Yang2021,Xu2021,Sun2021}, both of which are relatively common. More unusual spin structures were also observed such as a topological multi-$\vec{k}$ structure in CeAlGe~\cite{Puphal2020a}, a spiral order in SmAlSi~\cite{Yao2022}, and a helical incommensurate spin density wave (SDW) in NdAlSi~\cite{Gaudet2021}. In particular, for NdAlSi, its helical magnetism was shown to be stabilized by bond-oriented Dzyaloshinskii-Moriya (DM) interaction predicted to arise from Weyl-mediated Ruderman-Kittel-Kasuya-Yosida (RKKY) coupling, owing to the presence of itinerant Weyl electrons, local magnetic moments, and broken inversion symmetry~\cite{Gaudet2021,Nikolic2020a}.\

Despite the comprehensive characterization of Weyl-mediated magnetism in NdAlSi \cite{Gaudet2021} and some studies on NdAlGe \cite{Zhao2022}, their transport properties such as anomalous Hall effect (AHE) remain unexplored. AHE has been extensively investigated in FM Weyl semimetals where the intrinsic Berry curvature may contribute to pronounced AHE~\cite{Liu2018b,Wang2018b,Burkov2014a}, but recently it has been shown that Berry curvature is not always the dominant source of AHE in FM Weyl semimetals, and extrinsic disorders can also play a major role~\cite{Yang2020d}. With a chiral magnetism established in NdAlSi, NdAl$X$ provides a unique system to study AHE in helimagnetic Weyl semimetals. In this work, we aim to first establish the magnetic structure of NdAlGe, which is not obvious (not necessarily the same as NdAlSi) considering the behavior in other materials in $R$Al$X$ family \cite{Hodovanets2018,Puphal2019a,Yang2021,Yang2020d}, and study the AHE of NdAl$X$ with a focus on NdAlGe to understand the interplay among topology, magnetism, and electrical transport.

We investigate the magnetism and electrical transport of NdAlGe with SQUID magnetometry, heat capacity measurements, neutron scattering experiments, resistivity measurements, and DFT calculations. Similar to NdAlSi, we found a high temperature ($5.1\ \text{K} < T < 6.8$ K) helical incommensurate SDW in NdAlGe characterized by a multi-$\bold{k}$ structure with ordering vectors $\bold{k}_{\text{AFM1}}=(2/3+\delta(T),2/3+\delta(T),0)$, $\bold{k}_{\text{AFM2}}=(1/3-\delta(T),1/3-\delta(T),0)$, and $\bold{k}_{\text{FM}}=(3\delta(T),3\delta(T),0)$, which evolves to a commensurate ($\delta~=~0$) helical ferrimagnetic state at low temperatures ($T<5.1$ K). In this state, the small-angle neutron scattering (SANS) can be modeled by an anisotropic Lorentzian-squared function, which signifies stripes of ferrimagnetic domains with real space characteristic length scales of 18(5)~nm and 72(8)~nm parallel and perpendicular to the [110] ordering vector direction, respectively. Surprisingly, we found AHE responses as Hall resistivity plateaus under a magnetic field in NdAlGe but not in NdAlSi, despite that both materials show clear plateaus in their magnetization. Furthermore, as the field increases, the AHE in NdAlGe shows a transition from intrinsic AHE in the ferrimagnetic phase to an extrinsic AHE in the polarized FM phase. Finally, we calculate the electronic band structure, Weyl-nodes, and Fermi surface of NdAlGe. We also calculated the anomalous Hall conductivity of NdAlGe, which shows a reasonable agreement with the observed intrinsic AHE. Our findings of helical magnetism and two regimes of AHE in NdAlGe suggest that Weyl-mediated magnetism is robust in these materials provided that the itinerant Weyl electrons and nesting Fermi pockets are intact, while AHE can be largely modified by extrinsic disorders and spin fluctuations despite significant intrinsic Berry curvature.


\section{Methods}\label{methods}
Single crystals of NdAlGe and NdAlSi were grown by a self-flux method. The starting materials are elemental chunks of Nd, Al, and Ge, weighed in a ratio of Nd:Al:Ge$=$1:10:1 or 1:15:1, and mixed in an alumina crucible. Single crystals made with a 1:10:1 recipe were used in this study unless specified. NdAlSi single crystals were grown with the 1:10:1 ratio. The crucibles were sealed in an evacuated quartz tube, heated up to 1050$^\circ$C at 3$^\circ$C/min, dwelt for 12 hours, cooled down to 700$^\circ$C at 0.1$^\circ$C/min, and dwelt for another 12 hours.  After the heating sequence, the tube was centrifuged to remove the excess Al flux. Plate-like single crystals were found isolated and attached to the bottom of crucibles. X-ray diffraction (PXRD) measurement was performed with a Bruker D8 ECO instrument with a copper x-ray source (Cu K$\alpha$) and a one-dimensional LINXEYE-XE detector at room temperature. Rietveld refinement on the PXRD patterns was performed using the FullProf suite \cite{Rodriguez-Carvajal1993}. The elemental analysis of NdAlSi/Ge was determined by Energy Dispersive X-ray Spectroscopy (EDX), carried out with FEI Scios, operated at an acceleration voltage of 20 kV and a current of 0.4 nA.

Electrical resistivity and heat capacity were measured in a Quantum Design Physical Property Measurement System (PPMS) Dynacool with the standard four-probe technique and relaxation time method, respectively. DC magnetization experiments
were performed on the vibrating sample magnetometer in a
Quantum Design MPMS3.\

We performed neutron diffraction using both the HB-1A thermal triple-axis spectrometer at ORNL and the cold triple-axis spectrometer SPINS at NIST. The (HHL) scattering plane of NdAlGe was probed for both experiments at a base temperature of $\sim$1.5~K. 14.5~meV incident neutrons filtered with pyrolytic graphite (PG) were used on HB-1A. For SPINS, we used 3.7~meV incident neutrons with cooled Be filters employed both before and after the sample.\


The HFIR GP-SANS instrument was utilized to probe the SANS of NdAlGe with a base temperature of $\sim$2~K. Two SANS configurations were used; 1) Uncollimated 12~$\angstrom$ incident neutrons were used with the scattered neutrons detected at a distance of 19~m away from the sample, 2) 4.75~$\angstrom$ neutrons collimated by 3 guides were incident to the sample and detected at a distance of 8~m from it. A circular aperture with a diameter of 8~mm was placed at the sample position. A flat-plate sample was aligned such that the $[001]$ axis is parallel to the neutron beam. SANS patterns were collected using an 11~T horizontal magnet applied both parallel ($\bold{H}\parallel [001]$) and perpendicular ($\bold{H}\parallel[110]$) to the incident neutron beam. Error bars associated with all neutron diffraction intensities reported in this work correspond to 1 standard deviation.\

We performed electronic structure calculations within the framework of density functional theory (DFT) based on the projected augmented wave (PAW) method as implemented in the Vienna ab-initio simulation package (VASP) \cite{Hohenberg1964,Kresse1996,Kresse1999}. Generalized gradient approximation (GGA) \cite{Perdew1996} was used to include exchange-correlation effects and spin-orbit coupling (SOC) was added self-consistently. We added an on-site Coulomb interaction with $U_{\text{eff}}=8$ eV for the Nd $f$ electrons within the GGA+U scheme \cite{Anisimov1991,Anisimov1997a} to include strong electron-correlation effects. We considered the kinetic energy cut-off of 450~eV for the plane-wave basis set and used $\Gamma$-centered $11\times11\times11$ $k$-mesh \cite{Monkhorst1976} for bulk Brillouin zone sampling. The tolerance of the electronic energy minimization was set to $10^{-6}$ eV. We generated material-specific tight-binding Hamiltonian using the VASP2WANNIER90 interface~\cite{Pizzi2020}. We included Nd $d$, $f$, Al $s$, $p$ and Ge $s$, $p$ orbitals in construction of the Wannier functions. The topological properties were calculated using the WannierTools package~\cite{Wu2018a}.\

\section{Results}
Table \ref{tab:table1} summarizes the main properties studied in this paper, including helical magnetism and anomalous Hall effect. When comparing NdAlGe and NdAlSi together, we find that their magnetic properties are similar to each other, but their transport properties are quite different. Incommensurate and commensurate magnetic orders are found in both materials at close temperatures, and their refined spin structures are also similar to each other~\cite{Gaudet2021}. However, when it comes down to the transport properties, the residual resistivity ratio (RRR) of NdAlGe is 3 times smaller than that of NdAlSi, which suggests a higher disorder level in NdAlGe. The lower RRR in NdAlGe is also manifested in the single-band analysis, which shows a higher hole concentration but a much lower hole mobility compared to NdAlSi \footnote{We note that a single-band analysis does not give us quantitatively accurate carrier density and mobility of all carriers, but it gives us a qualitative picture of the dominant carrier and helps us understand the difference between NdAlSi and NdAlGe.}. As we will see, in spite of similar Fermi surfaces, quantum oscillations are not observed in NdAlGe, but are pronounced in NdAlSi~\cite{Gaudet2021} at the same temperatures and magnetic fields; this distinction also suggests a shorter mean free path and lower mobility in NdAlGe. More interestingly, both intrinsic and extrinsic AHE were observed in NdAlGe as magnetic field increases, but no clear sign of AHE was found in NdAlSi. In the following sections, we will present and interpret the structural, magnetic, and electrical transport properties of NdAlGe in detail.\

\begin{table}[t]
\caption{\label{tab:table1}%
Summary of magnetic and transport properties of NdAlSi and NdAlGe, including the onset of incommensurate order ($T_\text{inc}$) and commensurate order ($T_\text{com}$), saturated moment $M_{\text{sat}}$, residual resistivity ratio (RRR, defined as $\rho(300\text{K})/\rho(2\text{K})$, single-band carrier concentration ($n_h$, in both materials hole is the dominant carrier) and mobility ($\mu_h$), and anomalous Hall effect (AHE).
}
\begin{ruledtabular}
\begin{tabular}{c|cc}
 &
NdAlGe &
NdAlSi \cite{Gaudet2021} \\
\colrule
$T_{\text{inc}}$  & 6.8(2) K   & 7.2 K\\
$T_{\text{com}}$  & 5.1(1) K   & 3.3 K\\
$M_{\text{sat}}$ at 2 K, 6 T & 2.8(1) $\mu_{\text{B}}$  & 2.9 $\mu_{\text{B}}$\\
RRR &  2.0(1)  & 6.0\\
$n_h$ at 2 K & $1.06 \times 10^{21}$ cm$^{-3}$  & $6.66 \times 10^{19}$ cm$^{-3}$\footnote{\label{note1}See Appendix \ref{app:app_rhoxx_rho_yx_comp}.}\\
$\mu_h$ at 2 K & 134 cm$^{2}$V$^{-1}$s$^{-1}$  & 11008 cm$^{2}$V$^{-1}$s$^{-1}$\footnoteref{note1}\\
\multirow{2}{*}{AHE}& duu state: intrinsic &\multirow{2}{*}{} No clear Hall\\
    & FM state: extrinsic & resistivity plateau\footnote{\label{note2}See Appendix \ref{app:app_rhoyx_MH_NdAlSi}.}\\
\end{tabular}
\end{ruledtabular}
\end{table}

\subsection{Crystal structure and disorder in NdAlGe}\label{xrd}

\begin{figure}[t]
    \includegraphics[width=\columnwidth]{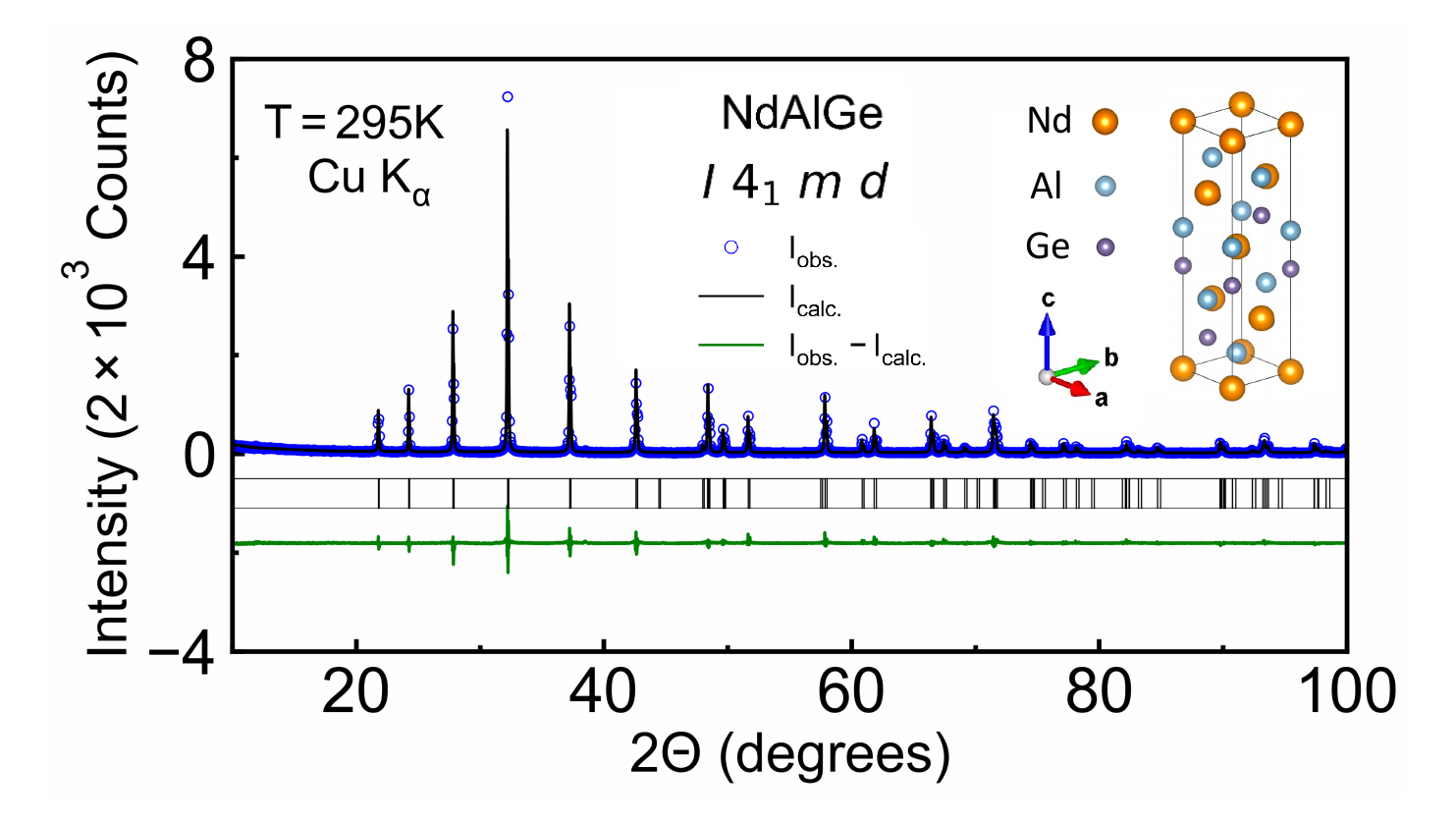}
    \centering
    \caption{Powder x-ray diffraction pattern and Rietveld refinement of NdAlGe. The occupancy of each atom was refined. The crystal structure of NdAlGe is presented on the top right inset.}
    \label{fig:xrd}
\end{figure}

\begin{table}[b]
\caption{\label{tab:table2}%
EDX measurements of NdAlSi and NdAlGe. The occupancy was normalized by that of Al. The uncertainty was defined by the standard deviation of all measurements, which were taken from 2-3 different regions of several crystals for each material.
}
\begin{tabularx}{1\columnwidth} { 
   >{\centering\arraybackslash}X 
  | >{\centering\arraybackslash}X
  || >{\centering\arraybackslash}X
   >{\centering\arraybackslash}X}
 \hline
 \hline
 NdAlSi & Occupancy & NdAlGe & Occupancy \\
 \hline
Nd  & $0.99\ \pm$ 0.01 & Nd  & $0.94\ \pm$ 0.04\\
Al  & $1\ \pm$ 0.01 & Al  & $1\ \pm$ 0.07\\
Si  & $0.99\ \pm$ 0.01 & Ge  & $0.93\ \pm$ 0.04\\
\hline
\hline
\end{tabularx}
\end{table}

The inset of Fig.~\ref{fig:xrd} shows the crystal structure of NdAlGe, which belongs to the same $I$4$_1md$ space group as the archetypal Weyl semimetal TaAs~\cite{Huang2015e,Weng2015,lv_experimental_2015,yang_weyl_2015}. The combination of the noncentrosymmetric crystal structure and the collective magnetism hosted by Nd$^{3+}$ $f$-orbitals at low temperatures makes NdAlGe a double-symmetry-breaking Weyl semimetal~\cite{Chang2018}.
Based on previous second harmonic generation experiments across different $R$Al$X$ compounds, it is now clear that the $R$Al$X$ material family resides in the noncentrosymmetric space group $I$4$_1md$~\cite{Yang2020d,Yang2021,Gaudet2021,Yao2022}. We will then use the $I$4$_1md$ space group as our starting point for the nuclear structure refinement of NdAlGe.\

Another important point is the stoichiometry of $R$Al$X$ single crystals that may vary depending on the growth methods. In terms of the growth methods, it has been shown that single crystals grown by a floating-zone furnace typically has a stoichiometric ratio much closer to 1:1:1 compared to those grown by flux methods, which tend to be Al-rich and have vacancies on the Ge/Si sites \cite{Puphal2019a}. Furthermore, current literature seems to suggest the Ge variant is more prone to off-stoichiometry as compared to their Si analogue. For example, CeAlSi crystals grown by the flux method show little deviation from a 1:1:1 stoichiometric ratio~\cite{Yang2021}; however, for CeAlGe, the crystals grown by the flux method are predominantly Al-rich and have significant vacancies on the Ge sites, while those synthesized with a floating-zone growth show a stoichiometric ratio close to 1:1:1~\cite{Puphal2019a}.\

Considering $R$AlGe grown by flux methods are prone to be Al-rich, we have performed EDX measurements to confirm the stoichiometry in our NdAl$X$ crystals grown by flux method (Table \ref{tab:table2}). Our measurements show that NdAlSi crystals are close to a 1:1:1 stoichiometry, while NdAlGe crystals are predominantly Al-rich and show larger variations in the stoichiometry. We also refined the atomic occupancy for both materials by Rietveld refinement, and the results show a similar trend (see Appendix \ref{xrd_Refinement}). Our characterizations are consistent and suggest that, compared to NdAlSi, the flux-grown NdAlGe crystals show variation in their atomic compositions and have a higher level of disorder.


\subsection{Resistivity, magnetic susceptibility, and heat capacity}

\begin{figure}[t]
    \includegraphics[width=\columnwidth]{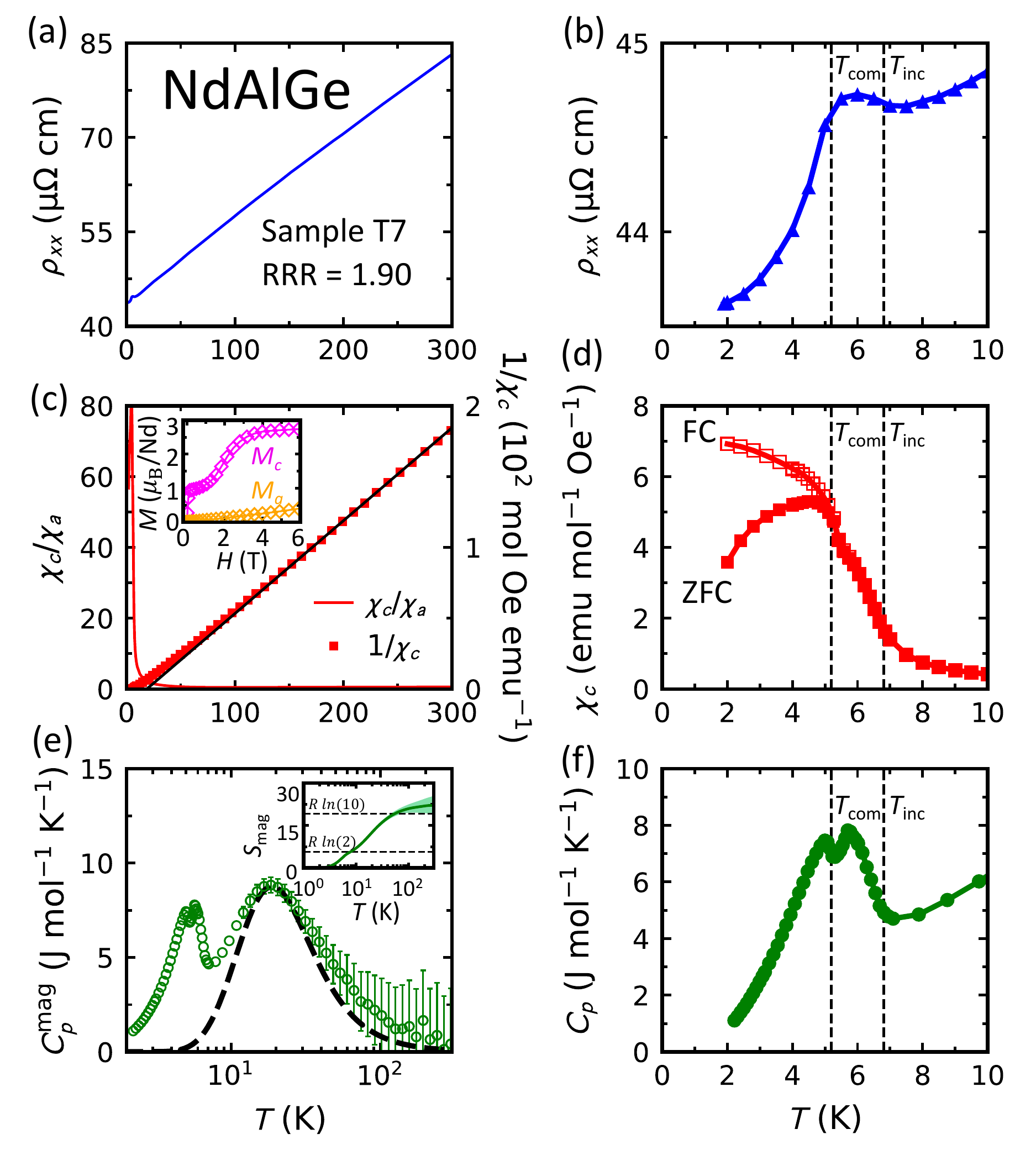}
    \centering
    \caption{(a,b) Resistivity as a function of temperature, plotted from 0 to 300~K and 0 to 10~K, respectively. $T_\text{com}=5.1$ K and $T_\text{inc}=6.8$ K respectively indicates the transition temperature of the commensurate and incommensurate orders. (c) The ratio of magnetic susceptibility, which was measured with field $H=100$ Oe along the c-axis ($\chi_c$), to that measured with the same field along the a-axis ($\chi_a$). Both $\chi_c$ and $\chi_a$ are measured after cooling down to 2~K in zero magnetic field (ZFC). $1/\chi_c$ is also plotted in the same panel, and the black line shows the result of a Curie-Weiss fit to the data above 150 K. The inset shows the magnetization of NdAlGe measured at $T~=~2$~K with the magnetic field applied along the c-axis ($M_c$) and a-axis ($M_a$). (d) $\chi_c$ measured while the sample is cooled under field $H=100$ Oe (FC) and $\chi_c$ measured under ZFC. (e) Temperature dependence of the magnetic heat capacity $C_p^{\text{mag}}$ of NdAlGe, and the error bars account for the mass uncertainty (4$\%$) of the sample. The inset shows the magnetic entropy $S_{\text{mag}}$ and the shaded area represents the uncertainty.. $C_p^{\text{mag}}$ was obtained by subtracting the heat capacity of LaAlGe from the heat capacity of NdAlGe. The dashed black line in the main panel is the predicted "Schottky-like" $C_p^{\text{mag}}$ anomaly calculated assuming that the $(2J+1)$ spin-orbit levels of Nd$^{3+}$ \textcolor{red}{($J~=~9/2$)} are split by crystal electric field (CEF) effects into a doublet ground state, \textcolor{red}{3} excited doublets at 4~meV, and another excited doublet at 9~meV. (f) The total heat capacity below $T=10$ K.}
    \label{Fig2_com_inc_Tdep}
\end{figure}

Figure \ref{Fig2_com_inc_Tdep}(a,b) shows a typical resistivity ($\rho_{xx}$) curve obtained for our NdAlGe crystals at $T=2-300$ K and $T=2-10$ K, respectively. From the magnitude of $\rho_{xx}$ at $T=300$ K and $T=2$ K, we calculate the residual resistivity ratio $\text{RRR}=\frac{\rho_{xx}(300\ \text{K})}{\rho_{xx}(2\ \text{K})}$ to be 1.9, much lower than the RRR of NdAlSi, which is 6.0 (see Appendix \ref{app:app_rhoxx_rho_yx_comp}). A lower RRR usually suggests the disorder level is higher so that the resistivity is anchored at a higher value near zero temperature. We attribute the lower RRR for NdAlGe to its off-stoichiometry characterized in the previous section.

We measured the magnetic susceptibility of NdAlGe ($\chi$) with a magnetic field applied along the c-axis ($\chi_c$) and the a-axis ($\chi_a$). The ratio $\chi_c/\chi_a$ is plotted as a solid line in Fig.~\ref{Fig2_com_inc_Tdep}(c) (left y-axis), while $1/\chi_c$ is plotted as squares in the same plot (right y-axis). We also plotted the temperature dependence of $\chi_c$ in Fig.~\ref{Fig2_com_inc_Tdep}(d). At high temperatures, the magnetic susceptibility shows isotropic paramagnetic spins with a Curie-Weiss temperature of 18.0(7) K and an effective magnetic moment of 3.5(2) $~\mu_{B}$. Upon cooling, similar to NdAlSi~\cite{Gaudet2021}, an out-of-plane anisotropy gradually develops such that the ratio $\chi_c/\chi_a$ reaches as high as 80 at $T=2$~K. The out-of-plane anisotropy is also visible in the low-temperature in-field magnetization of NdAlGe (inset of Fig.~\ref{Fig2_com_inc_Tdep}(c)) where the magnetization along the $c$-axis reaches saturation near 3~T at a value of 2.8(1)$\mu_B/Nd$, while the magnetization along the $a$-axis is still unsaturated and weak at 6~T.\

The main panel of Fig.~\ref{Fig2_com_inc_Tdep}(e) shows the magnetic contribution of the heat capacity $C_p^{\text{mag}}$ of NdAlGe, which was obtained by subtracting the heat capacity of the non-magnetic analogue compound LaAlGe (see Appendix \ref{app:app_HC}). A magnetic Schottky-like anomaly centered at 18(1)~K is visible in $C_p^{\text{mag}}$. Consistent with the magnetic entropy (inset of Fig. \ref{Fig2_com_inc_Tdep}(e)), this high-temperature anomaly can be reproduced with an Nd$^{3+}$ single-ion energy scheme comprised of a doublet ground state separated by \textcolor{red}{4} excited doublets between 3 to 9~meV. At lower temperatures, $C_p^{\text{mag}}$ shows two additional anomalies at $T_{\text{com}}$~=~5.1(1)~K and $T_{\text{inc}}$~=~6.8(2)~K, which originate from the collective magnetism of the Nd$^{3+}$ moments (Fig.~\ref{Fig2_com_inc_Tdep}(f)). The presence of two anomalies is also observed in NdAlSi where the phase transition at $T_{\text{inc}}$ signifies the onset of an incommensurate modulated spin density wave that transitions into a commensurate ferrimagnetic state below $T_{\text{com}}$.\ 

The incommensurate and commensurate magnetic phase transitions in NdAlSi were both observed to impact its electric transport and bulk thermodynamic properties. More specifically, the onset of the commensurate order in NdAlSi can be seen as discontinuity occurring at $T_{\text{com}}$ in $\rho_{xx}(T)$, $\chi(T)$, and $C_p(T)$. For the incommensurate order, however, it is less obvious, but NdAlSi shows a drop in $\rho_{xx}(T)$, an arguable change of slope in $\chi(T)$, and a clear peak in $C_p(T)$ at $T_{\text{inc}}$~\cite{Gaudet2021}. For comparison, we also looked for similar effects in NdAlGe. In Fig.~\ref{Fig2_com_inc_Tdep}(b,d,f), we again see clear features at $T_{\text{com}}$ as a drop in $\rho_{xx}$, a split of FC and ZFC data in $\chi_c$, and a peak in $C_p$. Anomalies associated with the incommensurate order of NdAlGe are still subtle in $\rho_{xx}(T)$ and $\chi(T)$, where a mild upturn and a mild change of slope are observed at $T_{\text{inc}}$, respectively. In the heat capacity data, however, there is a peak that starts at $T_\text{inc}$ and one can argue the presence of two transitions in $C_p(T)$. The low temperature $C_p$ peaks of NdAlGe are broader than the ones observed for NdAlSi, which have sharp discontinuities occurring exactly at $T_{\text{com}}$ and $T_{\text{inc}}$. This observation suggests a similar magnetic phase diagram for both NdAlSi and NdAlGe, but with more disorder in NdAlGe.\

\subsection{Neutron diffraction}

To gain insights into the collective magnetism of NdAlGe, we have performed single-crystal neutron diffraction to determine its temperature-dependent spin structure. Below $T_{\text{inc}}$, we found incommensurate magnetism in NdAlGe that is characterized by strong magnetic Bragg peaks ($\bold{Q}_{\text{mag}}$) indexed with an ordering vector $\bold{k}_{\text{AFM1}}=(2/3+\delta(T),2/3+\delta(T),0)$, as well as weaker Bragg peaks indexed with $\bold{k}_{\text{AFM2}}=(1/3-\delta(T),1/3-\delta(T),0)$.\

\begin{figure}[t]
    \includegraphics[width=\columnwidth]{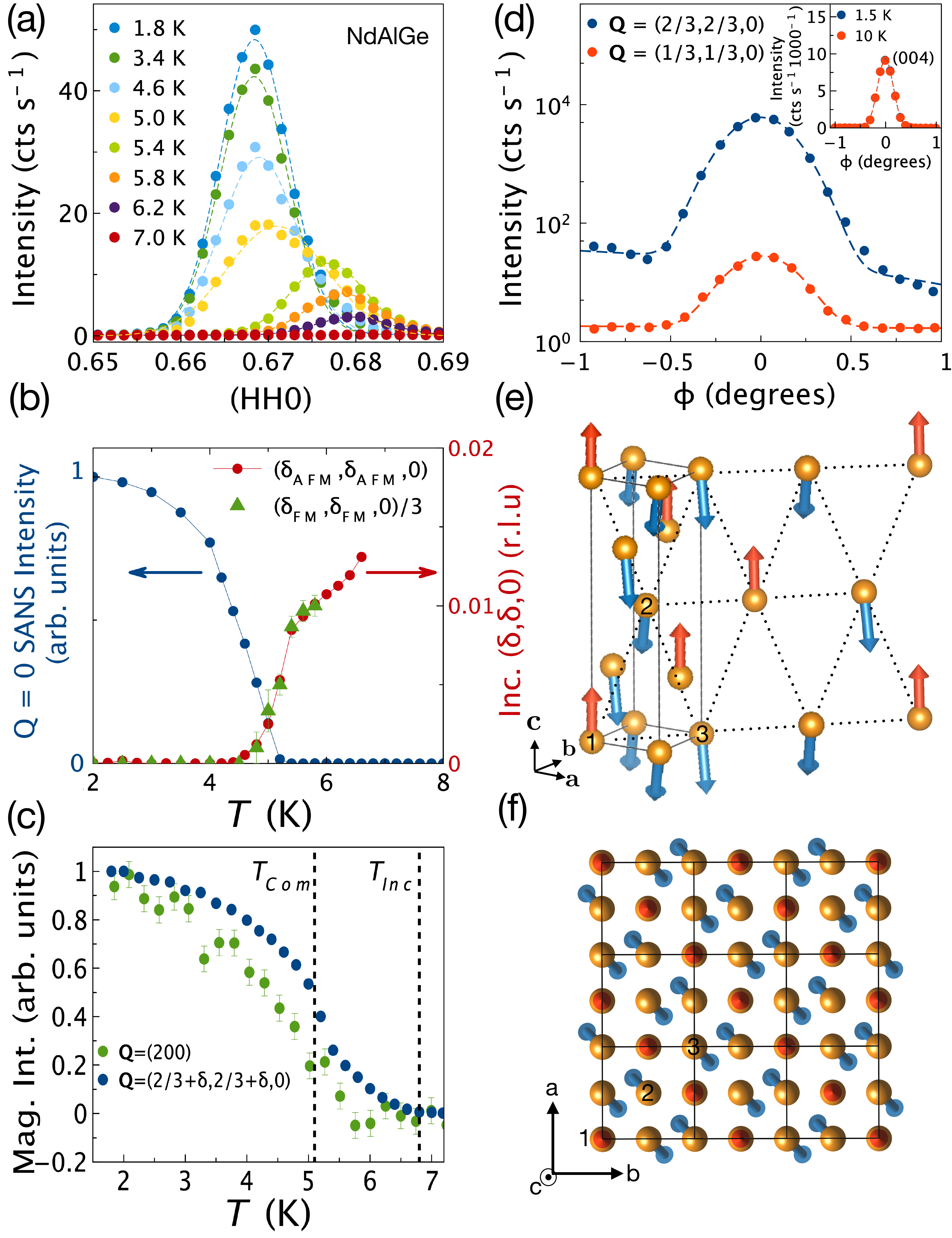}
    \centering
    \caption{(a) Neutron diffraction scans collected along the reciprocal (HH0) space direction at various temperatures between 1.5~K to 8~K centered around $\bold{Q}~=~(2/3,2/3,0)$. (b) The temperature dependence of the $(\delta(T),\delta(T),0)$ incommensurability of the $\bold{k}_{\text{AFM1}}=(2/3,2/3,0)$ SDW is plotted in red. The green curve shows 1/3 of the $(\delta_{\text{FM}}(T),\delta_{\text{FM}}(T),0)$ incommensurability of the ferromagnetic component whereas the blue curve is the total SANS collected using the $12~\angstrom$ data. (c) The order parameter of $\bold{Q}=(200)$ and $\bold{Q}~=~(2/3+\delta(T),2/3+\delta(T),0)$ Bragg peaks. (d) The main panel shows the rocking scans at $\bold{Q}~=~(2/3,2/3,0)$ and $\bold{Q}~=~(1/3,1/3,0)$ for $T=1.5~K$. We note that the source of the elevated background for the $\bold{Q}~=~(2/3,2/3,0)$ rocking scan comes from proximity to an Al Bragg peak. The inset of panel (d) shows the $\bold{Q}~=~(0,0,4)$ rocking scans collected for both $T=10~K$ and $T=1.5~K$. (e) and (f) show the refined spin structure for the magnetic commensurate phase of NdAlGe. Red (blue) arrows are used to represent the up (down) spins. The tilting of the spins within the $ab$ plane was amplified by a factor of 4 to allow for better visualization.}
    \label{BulkDiff}
\end{figure}

As shown in Fig.~\ref{BulkDiff}(a), we determined the incommensurability $\delta(T)$ by tracking the temperature dependence of the $\bold{Q}_{\text{mag}}=(2/3+\delta(T),2/3+\delta(T),0)$ peak center observed in an $(HH0)$ scan. The temperature dependence of $\delta(T)$ is plotted in Fig.~\ref{BulkDiff}(b) where a mild change of $\delta(T)$ between $T_{\text{com}} < T < T_{\text{inc}}$ is observed, but a transition to commensurate magnetism ($\delta~=~0$) arises for $T < T_{\text{com}}$. For comparison, we note that Fig.~\ref{BulkDiff}(b) also includes data from our SANS analysis, which will be presented in the next section. The order parameter of the $\bold{Q}_{\text{mag}}=(2/3+\delta(T),2/3+\delta(T),0)$ peak (Fig.~\ref{BulkDiff}(c)) correlates with $T_{\text{inc}}$.\ 

In addition to the antiferromagnetic $\bold{k}_{\text{AFM1}}$ and $\bold{k}_{\text{AFM2}}$, we also observed ferromagnetism in NdAlGe. To prove this, we acquired an order parameter at $\bold{Q}=(200)$ (Fig.~\ref{BulkDiff}(c)), which shows it onsets slightly above $T_{\text{com}}$. In the next section, we will see that ferromagnetism actually onsets exactly at $T_{\text{com}}$. The fact that we see magnetic intensity at $\bold{Q}=(200)$ above $T_{\text{com}}$ is from the onset of an incommensurate $\bold{k}=(\delta_{\text{FM}}(T),\delta_{\text{FM}}(T),0)$ wave that is a precursor to the $\delta_{\text{FM}}(T)=0$ ferromagnetism component. Our neutron diffraction experiment simply could not resolve this incommensurability, but we could do so using small-angle neutron scattering presented in the next section. \

The commensurate magnetic phase of NdAlGe is thus described by a multi-$\bold{k}$ spin structure including two different antiferromagnetic components ($\bold{k}_{\text{AFM1}}$ and $\bold{k}_{\text{AFM2}}$), as well as a ferromagnetic component ($\bold{k}=\bold{0}$). The diffraction pattern of NdAlGe is practically identical to the one observed in NdAlSi~\cite{Gaudet2021}. The possible magnetic basis vectors describing this spin structure were obtained by symmetry analysis and consist of the $xyz$ components of a SDW propagating along the [110] or [1$\Bar{1}$0] directions. The SDW can either have parallel or anti-parallel spins sitting on the primitive Nd$^{3+}$ sites at $\bold{r_1}~=~$(0,0,0) and $\bold{r_2}~=~$(1/2,0,1/4). Anti-parallel (parallel) spin components produce scattering at Bragg peaks indexed by the magnetic ordering vector $\bold{k}_{\text{AFM1}}=(2/3,2/3,0)$ ($\bold{k}_{\text{AFM2}}=(1/3,1/3,0)$). As seen in the main panel of Fig.~\ref{BulkDiff}(d), the Bragg peaks with $\bold{k}_{\text{AFM1}}=(2/3,2/3,0)$ have almost two orders of magnitude greater intensities than the $\bold{k}_{\text{AFM2}}=(1/3,1/3,0)$ ones. This indicates a dominant anti-parallel spin component for the SDW, which is augmented by a weak parallel one. The anti-parallel spin component was refined to an Ising one, while the parallel component to a weak in-plane spin canting that is transverse to the propagation of the SDW. The spin structure refinement of NdAlGe is shown in Appendix \ref{Refinement}.

The $\bold{k}=\bold{0}$ ferromagnetic part of the spin structure  was refined to a $c$-axis magnetized state. This is due to the fact that we did not observe magnetic Bragg intensity at nuclear-allowed $\bold{Q}=(0,0,L)$ Bragg positions (see top right inset of Fig.~\ref{BulkDiff}(d)), while we observed magnetic scattering at nuclear-allowed Bragg positions such as $\bold{Q}~=~(2,0,0)$ (Fig.~\ref{BulkDiff}(c)). The magnetization was refined to 0.9(1)$\mu_B/$Nd, which is consistent with the value of the low-field magnetization plateau reported in the inset of Fig.~\ref{Fig2_com_inc_Tdep}(c).

Finally, adding all spin components together, the spin structure of NdAlGe is an down-up-up (duu) ferrimagnetic SDW propagating along the [110] or [1$\Bar{1}$0] direction that is augmented by a weak in-plane chiral component lying transverse to its propagation (see sketch of the spin structures in Fig.~\ref{BulkDiff}(e,f)). Mostly pointing along the $c$-axis, the moment on each Nd sites was refined to 3.0(2)$\mu_\text{B}$ with an in-plane tilting angle of 3(1)$\degree$. The high temperature incommensurate spin structure is similar to the commensurate one, but convoluted with an amplitude-modulated wave that has a spatial wavelength of $\sim$ 35~nm. The resulting spin structure of NdAlGe is practically identical to the NdAlSi one, but we note that the incommensurate to commensurate phase transition in NdAlGe is much broader in temperatures than the one in NdAlSi~\cite{Gaudet2021}. For example, the 5~K (HH0) scan presented in Fig.~\ref{BulkDiff}(a) shows the presence of both a commensurate and an incommensurate peak, which signifies inhomogeneity within the NdAlGe crystal. This is consistent with a range of different critical temperatures $T_{\text{com}}$ coexisting within the same crystal of NdAlGe, which likely arises from a variation of the stoichiometry across the whole sample. Such a conclusion corroborates the fact that NdAlGe has more disorder than NdAlSi.

\subsection{Small-Angle neutron scattering (SANS)}

\begin{figure*}[t]
    \includegraphics[width=7in]{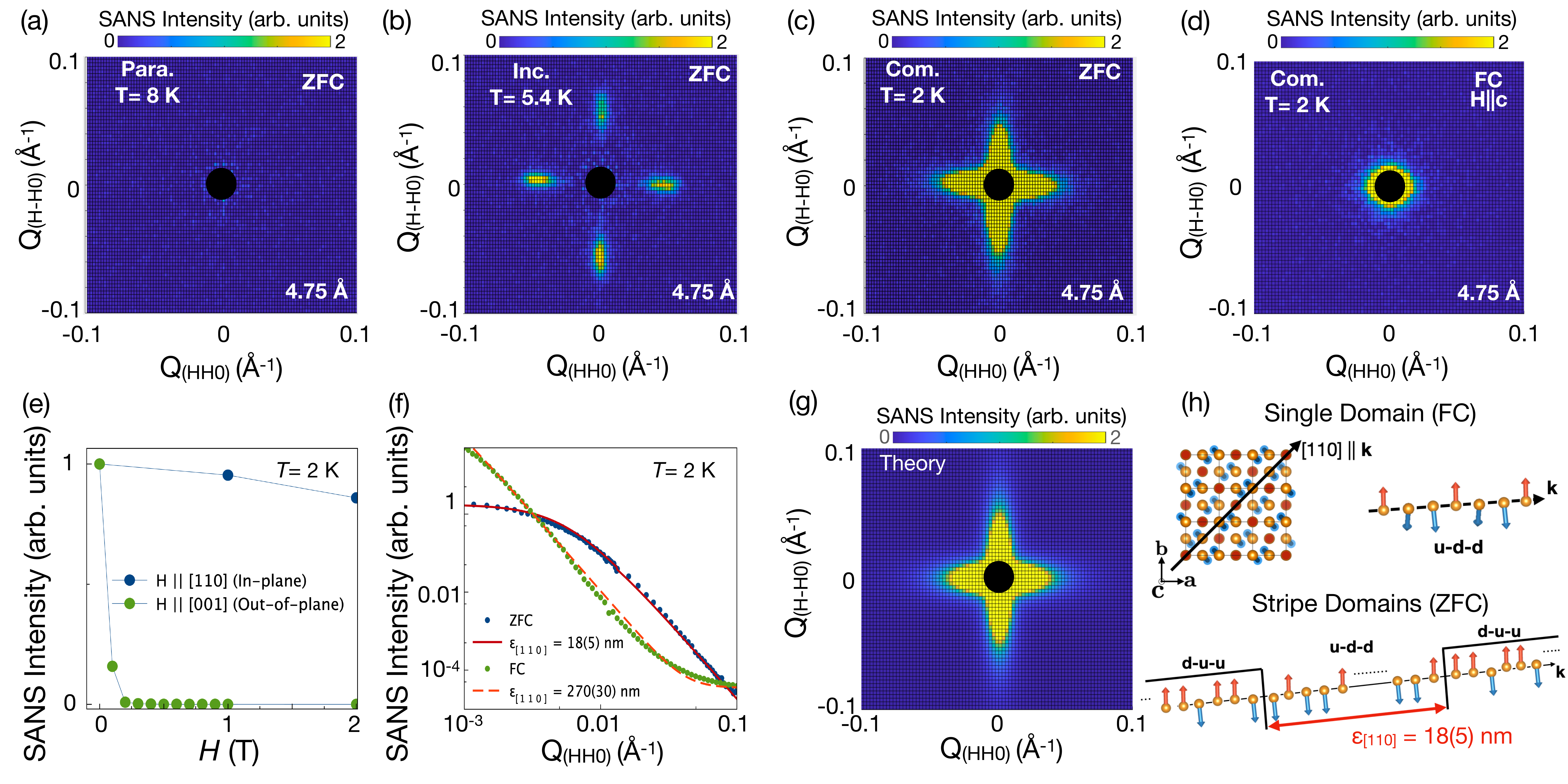}
    \centering
\caption{Panels (a),(b) and (c) correspond to the zero-field cooled (ZFC) $4.75~\angstrom$ SANS data respectively collected within the paramagnetic state (8~K), the incommensurate phase (5.4~K), and the commensurate phase (2~K). Panel (d) is the $4.75~\angstrom$ SANS data collected within the commensurate phase (2~K) using a field-cooled (FC) protocol (2~T). Panel (e) shows the total SANS intensity observed as a function of both an in-plane (blue) and out-of-plane (green) magnetic field using 4.75~$\angstrom$ neutrons at $T~=~2~K$. Panel (f) shows the $T~=~2~K$ total SANS scattering intensity as a function of the momentum transfer $\bold{Q}$ measured along the [H,H,0] direction for both ZFC and FC. This plot includes the SANS data collected within both the 4.75~$\angstrom$ and 12~$\angstrom$ configurations as well as their appropriate fit to a Lorentzian-squared function. Panel (g) is the calculated SANS pattern assuming an anisotropic Lorentzian-squared function with $\epsilon_{\parallel}$~=~18~nm and $\epsilon_{\perp}$~=~72~nm. Panel (h) shows a sketch of the 1D representation of the commensurate spin structure of NdAlGe (single domain). This 1D representation is also used to represent the stripe ferrimagnetic domains observed via SANS in a ZFC process.}
    \label{SANS}
\end{figure*}

So far, we have shown that the details of the magnetism of NdAlGe is impacted by the disorder. In order to characterize this further, we have performed a SANS experiment to probe its magnetized inhomogeneities on a spatial length scale of $\sim$ 1 to 500~nm.

We first collected field and temperature-dependent SANS data with the $\bold{c}$-axis parallel to the incident neutron beam so we could probe the in-plane scattering vectors. Representative data acquired with $4.75~\angstrom$ incident neutrons within the paramagnetic, incommensurate, and commensurate phase of NdAlGe are respectively shown in Fig.~\ref{SANS}(a,b,c). As expected, no coherent magnetic scattering is detected in the paramagnetic state. In the incommensurate phase, Bragg peaks at symmetry-related $\bold{Q}=(\delta_{\text{FM}}(T),\delta_{\text{FM}}(T),0)$ positions are observed corresponding to an SDW with a spatial modulation of 116(7)$~\angstrom$ at 5.4~K. This incommensurate SDW could not be resolved in our neutron diffraction experiment and is a precursor to the commensurate ferromagnetic spin component of NdAlGe occurring below $T_{\text{com}}$. The temperature dependence of the incommensurability $\delta_{\text{FM}}(T)$ is reported in Fig.~\ref{BulkDiff}(b). $\delta_{\text{FM}}(T)$ follows a $\delta(T)=\delta_{\text{FM}}(T)/3$ relationship indicating that the $\bold{k_{\text{FM}}}=(\delta_{\text{FM}}(T),\delta_{\text{FM}}(T),0)$ SDW is the third harmonic of the main $\bold{k}_{\text{AFM1}}=(2/3+\delta(T),2/3+\delta(T),0)$ wave. This is typical of incommensurate magnetism in rare-earth metallic systems where odd harmonics emerge from "squaring-up" of the main wave, which is expected upon cooling as the magnetization becomes more constant through the lattice~\cite{Taylor1971,Jensen1991}.

Within the commensurate phase, the incommensurate Bragg peaks disappear such that the SANS of NdAlGe is now centered at $|\bold{Q}|=0$ and is shaped like a cross extending along the $<$110$>$ directions (Fig.~\ref{SANS}(c)). This is different from the isotropic in-plane SANS pattern observed in the isostructural Weyl semimetal PrAlGe~\cite{Destraz2020}, and we will argue that the in-plane anisotropic cross pattern of NdAlGe arises from the finite size of the magnetic domains forming the multi-domains state. Using a field of 2~T, Fig.~\ref{SANS}(d) shows that field-cooling (FC) NdAlGe within its commensurate magnetic state significantly depletes the cross pattern.

To probe the temperature dependence of the cross pattern, we collected SANS data with $12~\angstrom$ incident neutrons, which exclude the magnetic incommensurate Bragg peaks such that the SANS scattering from the cross pattern can be isolated (see appendix~\ref{SANSApp}). The temperature dependence of the cross pattern extracted this way shows that it onsets below $T_{\text{com}}$ (blue circles in Fig.~\ref{BulkDiff}(b)) and is indeed a feature of the commensurate order.

We then studied the field evolution of the cross pattern by acquiring 4.75~$\angstrom$ SANS data for various in-plane and out-of-plane fields. As reported in Fig.~\ref{SANS}(e), an out-of-plane field of only 0.30(5)~T is enough to completely suppress the SANS intensity associated with the cross pattern, whereas a similar in-plane field strength does not significantly affect the scattering. This is consistent with the out-of-plane anisotropy of the bulk magnetization (inset of Fig.~\ref{Fig2_com_inc_Tdep}(c)) and shows that the cross is only observed when the time-reversal symmetric domains coexist (i.e. only when both up-down-down (udd) and duu domains are present).

We found that the SANS $\bold{Q}=\bold{0}$ cross pattern could be modeled using an anisotropic Lorentzian-squared function of the form $S(\bold{Q})=\frac{A}{((\epsilon_{\parallel}Q_{\parallel})^2+(\epsilon_{\perp}Q_{\perp})^2+1)^2}$, which is often used to phenomenologically describe the SANS of inhomogeneous magnetized systems such as spin glasses~\cite{aeppli1983spin,greedan1996frustrated,hellman1999long,muhlbauer2019magnetic}. In this expression, $A$ is a scale factor, while $\epsilon_{\parallel}$ and $\epsilon_{\perp}$ are the spatial correlation lengths of the ferrimagnetic domains parallel and perpendicular to the magnetic ordering vector direction [110], respectively. The momentum transfer $\bold{Q}$ is also expressed into a component that is either parallel ($Q_{\parallel}$) or perpendicular ($Q_{\perp}$) to the ordering vector. We note that there's also the presence of magnetic domains with ordering vector propagation along the [1$\Bar{1}$0] direction so we included a Lorentzian-squared function where $\epsilon_{\parallel}$ and $\epsilon_{\perp}$ are swapped. The combination of two Lorentzian-squared functions that represent the SDW along [110] and [1$\Bar{1}$0] accounts for the two branches of the cross pattern. The scattering function $S(\bold{Q})$ was then convoluted to the 2D resolution function ellipsoid of our SANS instrument and fitted to the 2D zero field SANS data of NdAlGe acquired at $T~=~2~K$ (Fig.~\ref{SANS}(c)). 

Following this procedure, we obtained $\epsilon_{\parallel}~=~18(5)$~nm and $\epsilon_{\perp}~=~72(8)$~nm. A quantitative comparison between the fit and the data is shown in Fig.~\ref{SANS}(f) for the momentum transfer along the [H,H,0] direction, while the resulting 2D fit is presented in Fig.~\ref{SANS}(g) and can be compared to Fig.~\ref{SANS}(c). Our result indicates the SANS observed within the commensurate phase of NdAlGe originates from ferrimagnetic stripes domains that have a shorter spatial length scale parallel to the SDW and a longer one perpendicular to the SDW. The ferrimagnetic stripes domains of NdAlGe are sketched in Fig.~\ref{SANS}(h). The anisotropic shape of the magnetic domains in NdAlGe may be a consequence of anisotropic exchange interactions, dipolar interactions, or Dzyaloshinskii-Moriya interactions. 

We associate the origin of the magnetic stripes in NdAlGe to the finite sizes of its bulk domains. This scenario is consistent with the observed field dependence of the $\bold{Q}=\bold{0}$ cross pattern. Indeed, contrary to an in-plane field, a field parallel to the $\bold{c}$-axis promotes one time-reversal domain over the other. In this case, a multi-domain sample is not preferred and the spatial dimensions of the energetically favoured domain then diverge. The same phenomenology explains the longer length scale observed in the low-temperature FC SANS data (Fig.~\ref{SANS}(d)), which is not expected if the cross pattern originates purely from domain wall scattering. A fit to the FC SANS data against the Lorentzian-squared scattering function (Fig.~\ref{SANS}(f)) shows that $\epsilon_{\parallel}=\epsilon_{\perp}=270(30)$~nm.

\subsection{Anomalous Hall effect}

\begin{figure}[t]
    \includegraphics[width=\columnwidth]{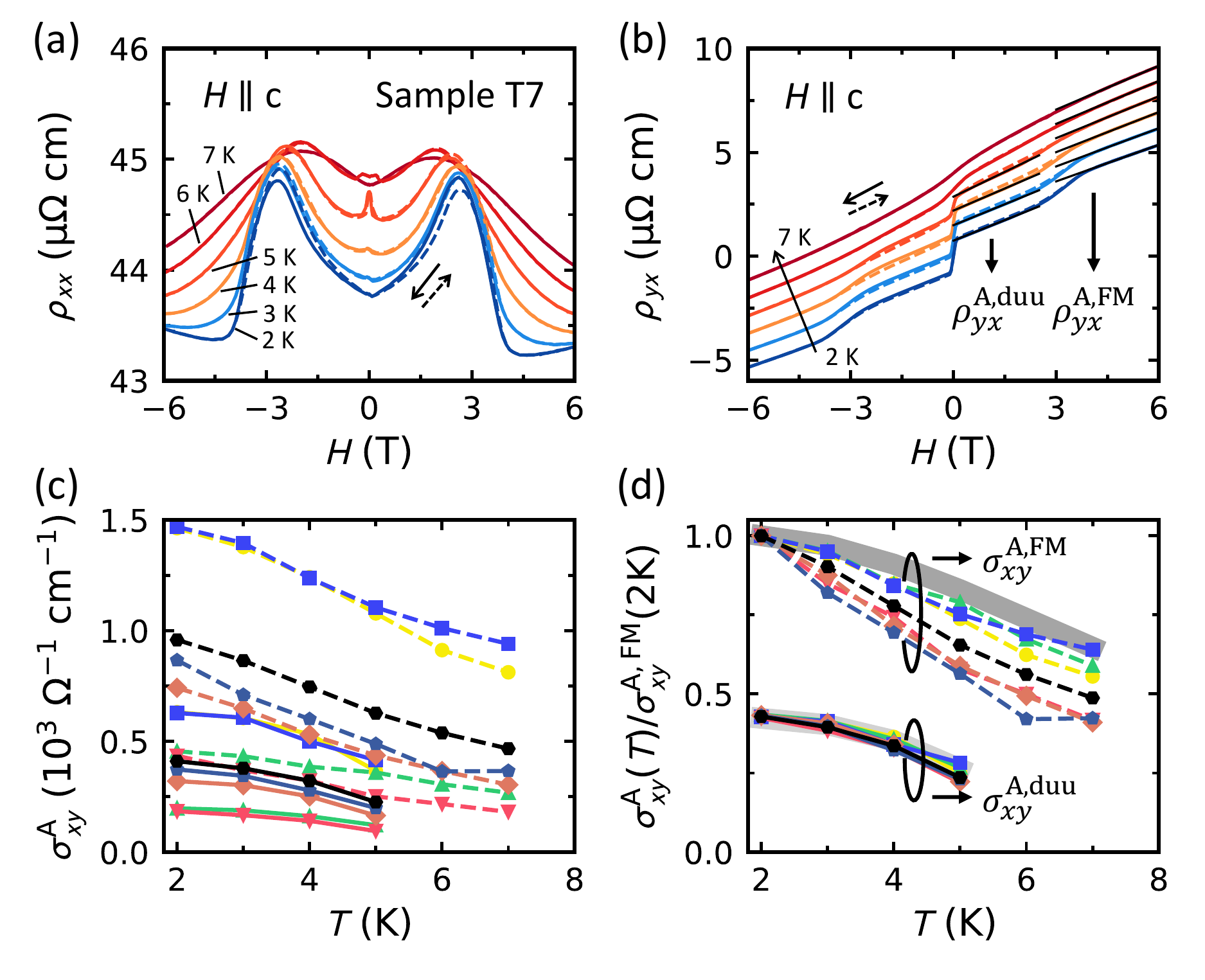}
    \centering
    \caption{(a) Resistivity $\rho_{xx}$ of the sample T7 as a function of external magnetic field $H$ below $T_{\text{inc}}$. The current is applied along the a-axis ($x$) and the field is along the c-axis ($z$). For each temperature, the data represented by a solid line are measured while the field is swept from $H=6$ T to $H=-6$ T, while the dashed line is recorded in the opposite field-sweeping direction. The same convention applies to panel (b). (b) Hall resistivity $\rho_{yx}(H)$ of the sample T7 collected at the same temperatures as in panel (a). The data taken at each temperature were antisymmetrized and shifted by $0.5$ $\mu \Omega\ \text{cm}$ from each other for visibility. The anomalous part of the Hall resistivity $\rho_{yx}^{\text{A,duu}}$ and $\rho_{yx}^{\text{A,FM}}$ are extracted from the y-intercept of a linear line fitted to the plateaus of duu ($0.2\ \text{T}<H<1$ T) and FM ($H>4$ T) states, respectively. (c) Anomalous Hall conductivity (AHC) as a function of temperature in the duu ($\sigma_{xy}^{\text{A,duu}}$, solid lines) and FM state ($\sigma_{xy}^{\text{A,FM}}$, dashed lines) of seven samples (shown as different symbols and colors). The data of sample T7 are plotted with black circles. (d) Normalized AHC plotted as a function of temperature. The light gray and dark gray stripes correspond to the re-scaled magnetization in the duu state ($M_\text{duu}$) and FM state ($M_\text{FM}$), which are extracted from the y-intercept of a linear line fitted to the plateaus of duu ($0.2\ \text{T}<H<1$ T) and FM ($H>4$ T) states, respectively (see Appendix \ref{app:app_MH}).}
    \label{fig:AHE}
\end{figure}

We now turn to the anomalous Hall effect (AHE) of NdAlGe, and show that its duu and FM states host different types of AHE. Fig.~\ref{fig:AHE}(a) shows the field dependence of the electrical resistivity $\rho_{xx}(H)$ of NdAlGe measured below the transition temperature $T_\text{inc}$. Near zero magnetic field, small features can be seen at $T=5$ K and 6 K; they correlate with the incommensurate phase as revealed in Fig. \ref{BulkDiff}(b). $\rho_{xx}(H)$ curves taken at opposite field-sweeping directions show a mild hysteresis below $H^* \simeq 3$ T, which is the transition field from the duu to the FM state (see $M_c(H)$ data in top inset of Fig.~\ref{Fig2_com_inc_Tdep}(c)). The hysteresis starts from $T=T_\text{com}$ and persists as the temperature decreases. Another feature related to the transition field $H^*$ is the local maximum of $\rho_{xx}(H)$. At $T=2$ K, $\rho_{xx}(H)$ first increases with the field in the duu state, peaks at $H^*$, and then starts to decrease as the system is going through a smooth transition from the duu to the FM state. Then, $\rho_{xx}(H)$ reaches a minimum at the end of the smooth transition, and finally starts to increase again when it is in the FM state. Such a non-monotonic behavior of $\rho_{xx}(H)$ can also be seen in some of the half-Heusler compounds such as DyPtBi, which shows multiple field-induced phase transitions and has relatively low mobility ($<1000$ cm$^{2}$V$^{-1}$s$^{-1}$)~\cite{Mun2016b}. We note that the local maximum at $H=H^*$ persists above $T_\text{com}$ and $T_\text{inc}$ where the transition between duu and FM states no longer exists; such non-monotonic magnetoresistance above $T_C$ has also been reported in DyPtBi and other half-Heusler compounds~\cite{Chen2020,Schindler2020,Pavlosiuk2020a}. We may qualitatively understand the behavior of $\rho_{xx}(H)$ in terms of the two-current model \cite{Fert1968,Blundell2001}, which suggests that the resistivity in the duu state ($\rho_{\text{duu}}$) is larger than within the FM state ($\rho_{\text{FM}}$). Assuming that both an up spin ($\rho_{\uparrow}$) and a down spin ($\rho_{\downarrow}$) contribute to the current in parallel, and also that $\rho_{\uparrow} \gg \rho_{\downarrow}$, we may then express $\rho_{\text{FM}} = (1/\rho_{\uparrow} + 1/\rho_{\downarrow})^{-1} \sim \rho_{\downarrow}$, which is a relatively low value. In the duu state, since the up and down spins admix as the spin wave propagates, both $\rho_{\uparrow}$ and $\rho_{\downarrow}$ approach an averaged value of the two. As a result, the resistivity in the duu state $\rho_{\text{duu}}$ has a significant contribution from $\rho_{\uparrow}$ and is thus larger than $\rho_{\text{FM}}$.

Figure.~\ref{fig:AHE}(b) shows the Hall resistivity $\rho_{yx}(H)$ measured at $T<T_\text{inc}$. Below $H\lesssim 3$ T, $\rho_{yx}(H)$ taken in opposite field-sweeps follow different traces, which resemble the loop-shaped Hall responses observed in other $R$Al$X$ materials \cite{Yang2021,Piva2023,He2023}; such a loop Hall effect might be a ubiquitous Weyl-mediated transport phenomenon in this group of Weyl semimetals. At $T=2$ K, there are two plateaus in $\rho_{yx}(H)$, both of which correlate with the magnetization plateaus observed in the duu and FM states. By fitting each plateau to a linear line, we extracted the anomalous part of $\rho_{yx}(H)$ in the duu state ($\rho^\text{A,duu}_{yx}$) and FM state ($\rho^\text{A,FM}_{yx}$) using the $y$-intercept of their respective fitting line. $\rho^\text{A,duu}_{yx}$ was extracted only for $T\leq T_\text{com}$ since beyond that temperature there is no duu state, but the spins are still polarized at high fields and high temperatures so $\rho^\text{A,FM}_{yx}$ was calculated up to $T\sim T_\text{inc}$. From the information extracted from Figs. \ref{fig:AHE}(a,b), we calculated the anomalous Hall conductivity (AHC) in the duu state ($\sigma^\text{A,duu}_{xy}$) and the FM state ($\sigma^\text{A,FM}_{xy}$) at each temperature using $\rho_0 \equiv \rho_{xx}(H=0)$ and $\rho^\text{A}_{yx}$ as $\sigma^\text{A}_{xy}=\frac{\rho^\text{A}_{yx}}{(\rho^\text{A}_{yx})^2 + \rho_0^2}$. The results are plotted in Fig. \ref{fig:AHE}(c) for seven different samples. Each sample is uniquely represented by a specific color and symbol; for example, the data of the sample T7 is plotted with black circles. The solid line is used for $\sigma^\text{A,duu}_{xy}$ and the dashed line is for $\sigma^\text{A,FM}_{xy}$. At first glance, the data are all over the place and it seems difficult to draw a clear conclusion.

However, since the resistivity is calculated from the resistance that depends on the geometric factors of each sample, the uncertainty in these sample-dependent factors may have contributed to the ``randomness" of the data in Fig.~\ref{fig:AHE}(c). To eliminate the trivial effect of geometric factors and extract the intrinsic properties of NdAlGe, we divided both $\sigma^\text{A,duu}_{xy}$ and $\sigma^\text{A,FM}_{xy}$ of each sample by its own $\sigma^\text{A,FM}_{xy}$ measured at 2 K ($\sigma^\text{A,FM}_{xy}(2\text{K}))$. Assuming $\rho_{xx} \gg \rho_{yx}$ \footnote{This assumption is valid in NdAlGe with $\rho_{xx}(H=0)>40\ \mu\Omega\text{cm}$ at 2 K and $\rho^\text{A}_{yx}\sim 1-2\ \mu\Omega\text{cm}$ at 2 K.}, we show that the normalized AHC $\sigma^\text{A}_{xy}(T)/\sigma^\text{A,FM}_{xy}(2\text{K})$ is free of geometric factors (the superscripts are omitted below for simplicity):
\begin{eqnarray*}
    \sigma_{xy}(T)/\sigma_{xy}(2\text{K}) &=& \frac{\frac{\rho_{yx}(T)}{\rho^2_{yx}(T)+\rho^2_{xx}(T)}}{\frac{\rho_{yx}(2\text{K})}{\rho^2_{yx}(2\text{K})+\rho^2_{xx}(2\text{K})}} \simeq \frac{\frac{\rho_{yx}(T)}{\rho^2_{xx}(T)}}{\frac{\rho_{yx}(2\text{K})}{\rho^2_{xx}(2\text{K})}}
    \\
    &=& \frac{\rho_{yx}(T)}{\rho_{yx}(2\text{K})}\frac{\rho^2_{xx}(2\text{K})}{\rho^2_{xx}(T)}.
\end{eqnarray*}
Since geometric factors do not depend on $T$, the geometric factor of $\rho_{yx}$ (involving sample thickness) and $\rho_{xx}$ (sample length, width, and thickness) are all eliminated in normalized AHC. We plotted the results in Fig. \ref{fig:AHE}(d) and found interesting characteristics of $\sigma^\text{A,duu}_{xy}$ and $\sigma^\text{A,FM}_{xy}$.

As shown in Fig. \ref{fig:AHE}(d), for $\sigma^\text{A,duu}_{xy}$, the normalized AHC of different samples all collapsed onto a single curve. In addition, as $T$ decreases, the normalized $\sigma^\text{A,duu}_{xy}$ scales with the magnetization of the duu state ($M_\text{duu}$, light gray stripe) such that it saturates at low temperatures. The convergence of the data from different samples, the scaling between $\sigma^\text{A,duu}_{xy}$ and $M_\text{duu}$, and the saturation of AHC at low $T$ are strong evidence for an intrinsic AHE \cite{Yang2020d,onoda2006intrinsic,miyasato2007crossover,Zeng2006,Checkelsky2008,Liu2018b}. 

However, in sharp contrast to $\sigma^\text{A,duu}_{xy}$, the normalized $\sigma^\text{A,FM}_{xy}$ of different samples in Fig. \ref{fig:AHE}(d) do not collapse but diverge into several curves. Indeed, as $T$ decreases, the normalized AHC does not follow the magnetization of the FM state ($M_{\text{FM}}$, dark gray stripe). Since the NdAlGe crystals grown by flux method are prone to Ge vacancies, we expect that extrinsic disorders vary in each crystal and govern the variance in $\sigma^\text{A,FM}_{xy}$. When the different disorder levels among all samples are taken into account, the convergence of $\sigma^\text{A,duu}_{xy}$ among them becomes quite nontrivial, and strongly suggests a robust intrinsic contribution to the AHE due to Berry curvature. The clear distinction between $\sigma^\text{A,duu}_{xy}$ and $\sigma^\text{A,FM}_{xy}$ marks two regimes of AHE in NdAlGe: an intrinsic AHE in the duu state and an extrinsic AHE in the FM state.

\section{Discussion}

\begin{figure}[t]
    \includegraphics[width=\columnwidth]{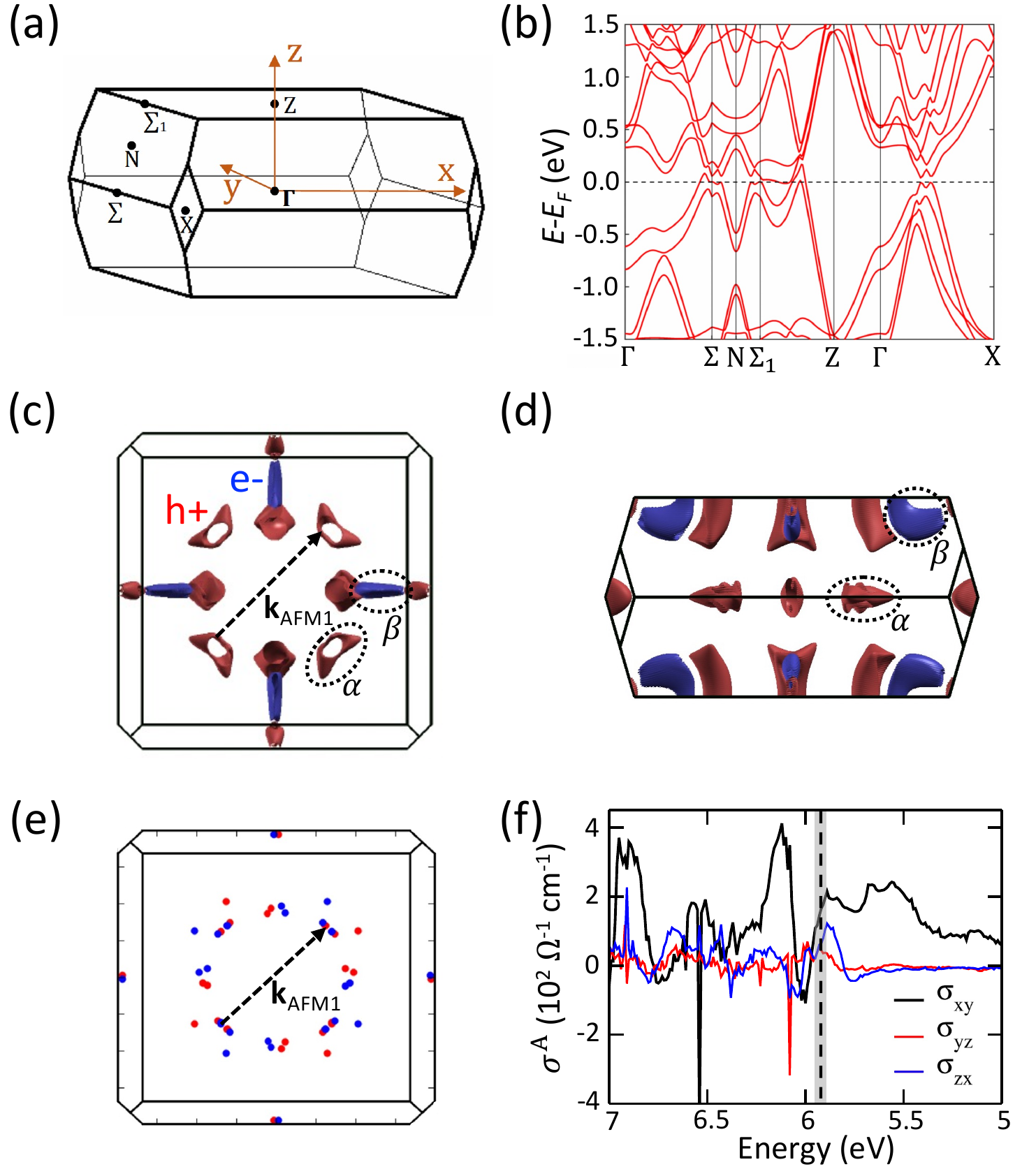}
    \centering
    \caption{(a) Brillouin zone of NdAlGe and high-symmetry $k$-points. (b) Band structure of NdAlGe. The dashed line marks the Fermi level calculated by DFT. (c) Fermi surfaces of NdAlGe. The blue pockets are electron pockets, while the red ones represent hole pockets. We further circle two pockets with dotted lines, and label the butterfly-shaped hole pocket as $\alpha$ and the elongated electron pocket as $\beta$ to facilitate our discussions. (d) Side view of the Fermi surfaces of NdAlGe. (e) The distribution of 56 Weyl nodes in the Brillouin zone found for the FM state of NdAlGe. (f) Anomalous Hall conductivity (AHC), calculated for different energy relative to the Fermi level determined by DFT ($E_F^{\text{DFT}}$, marked by the dashed line). The gray stripe near the dashed line marks the possible variation of Fermi levels in our NdAlGe samples; the variation range is $\pm 30$ meV based on the variation reported in other $R$Al$X$ materials \cite{Yang2021,Gaudet2021}.}
    \label{dft}
\end{figure}

To better understand and interpret the magnetism and AHE in NdAlGe, we calculate band structure, Fermi surface, Weyl nodes, and AHC due to Berry curvature. Fig.~\ref{dft}(a) shows the Brillouin zone and high-symmetry $k$-points; a $k$-path along these high-symmetry $k$-points was selected to plot the band structure of NdAlGe in Fig.~\ref{dft}(b). At the first glance, the band structure of NdAlGe does not look much different from that of NdAlSi~\cite{Gaudet2021}, but the similarities and differences are more visible when we look at the Fermi surface. From Fig.~\ref{dft}(c), we can see the butterfly-shaped hole pockets ($\alpha$ pockets) along the $\Gamma-X$ $k$-path, similar to the ones in NdAlSi near $\bold{Q} = (\pm \frac{1}{3}, \pm \frac{1}{3}, l)$~\cite{Gaudet2021}. These pockets fulfill the nesting condition for the incommensurate magnetic order to appear (see $\bold{k}_\text{AFM1}$ vector in Fig.~\ref{dft}(c)). Besides, when looking at both Fig. \ref{dft}(c) and (e) together, we find that the nesting $\alpha$ pockets are also Weyl-like and Weyl nodes near opposite $\alpha$ pockets along the diagonals are separated by the nesting wave vector. The inter-node scatterings between these Weyl nodes can provide the Weyl-mediated RKKY interactions and account for the chiral component in the duu ferrimagnetic order (helical magnetism)~\cite{Gaudet2021,Nikolic2020a}. The similarities between NdAlSi and NdAlGe in the nesting Fermi pockets and the distribution of Weyl nodes provide a reasonable explanation for their similar magnetic orders.

On the other hand, there are differences in the Fermi surfaces of NdAlSi and NdAlGe that distinguish their transport properties. The most pronounced difference lies in the diminished electron pockets in NdAlGe. In NdAlSi, in addition to the elongated electron pockets ($\beta$ pockets) along the $Z-\Sigma_1$ $k$-path at high $k_z$, an octagon-like network of electron pockets that extend to lower $k_z$ and connect the elongated pockets is also present \cite{Gaudet2021}. However, in NdAlGe, this network of electron pockets is diminished and only the $\beta$ pockets remain (Fig.~\ref{dft}(c,d)). Without this network, not only the number of electron carriers are reduced, but also the momentum dispersion of electrons is limited to a narrower range; both factors may explain the dominant role of hole carriers in NdAlGe.

In Fig.~\ref{dft}(f), we report the AHC contributed by intrinsic Berry curvature at different energies~\cite{Yao2004a}. At the Fermi level determined by our DFT calculations ($E_F^{\text{DFT}}$, indicated by the dashed line), $\sigma_{xy}^\text{A} \simeq 200\ \Omega^{-1}\text{cm}^{-1}$ agrees with both the sign and the order of magnitude of $\sigma_{xy}^\text{A,duu}$, which is $\simeq 400\ \Omega^{-1}\text{cm}^{-1}$ for sample T7. When taking possible variations of Fermi levels into account (indicated by the gray stripe), the calculated AHC remains positive and reaches a constant plateau towards the hole side (to the right of $E_F^{\text{DFT}}$). The constant value over a range of possible Fermi levels is consistent with the collapse of $\sigma_{xy}^\text{A,duu}$ data of all samples, which may have different Fermi levels. Although the calculation in Fig.~\ref{dft}(f) is done in the FM state, we expect it to reflect the AHC in the duu state because of the similarity in the net moment along $z$ in both states. For a more quantitative comparison, additional scaling analysis is required to determine the intrinsic AHC in $\sigma_{xy}^\text{A,duu}$. We tried to perform the scaling analysis proposed by Tian $et\ al.$~\cite{Tian2009a}; although the data points do follow the scaling ($\sigma_{xy}^\text{A,duu} \propto \sigma_{xx}^2$), the extracted intrinsic AHC seems unreasonably large, likely due to the narrow temperature range of the fitting, which is limited by $T_\text{com}$~\cite{ye2012temperature}. However, we argue that $\sigma_{xy}^\text{A,duu}$ should be dominated by intrinsic Berry curvature because 1) the collapse of data taken from samples of different disorders~\cite{Tian2009a,Yang2020d}, 2) the linear dependence of $\sigma_{xy}^\text{A,duu}$ on $M_\text{duu}$~\cite{Ye1999}, 3) the saturation of AHC at low temperatures \cite{miyasato2007crossover,Checkelsky2008}, 4) the conductivity $\sigma_{xx}$ falls in the regime where intrinsic AHE usually dominates~\cite{onoda2006intrinsic,miyasato2007crossover}, and 5) the reasonable agreement between $\sigma_{xy}^\text{A,duu}$ and $\sigma_{xy}$ calculated by DFT.

Previously, the transition from intrinsic to extrinsic AHE in $R$Al$X$ family was only observed among materials of different chemical compositions, and it was mainly driven by enhanced disorders~\cite{Yang2020d}. Here, we argue that spin fluctuations may play a key role in such a transition in NdAlGe. In ferromagnets, it has been proposed that carriers scattering in a fluctuating spin background may lead to a chirality-driven AHE~\cite{Ye1999,Tatara2002}, and a deviation of AHC from its scaling with $M$ was shown to be a manifestation of such behavior in experiments~\cite{Checkelsky2008}. In NdAlGe, $\sigma_{xy}^\text{A,FM}$ also deviates from $M_{\text{FM}}$ as $T$ increases (see  Fig.~\ref{fig:AHE}(d)). Besides, the magnetic fluctuations in the FM state seem to be strong, as suggested by the slow and smooth transition from the duu to the FM state, instead of the sharp and steep one in NdAlSi~\cite{Gaudet2021}. As a result, we infer that the enhanced spin fluctuations as the magnetic field increases, which possibly intensified with disorders, may be the key factor driving the transition from intrinsic to extrinsic AHE in NdAlGe. A complete description of $\sigma_{xy}^\text{A,FM}$, however, could be quite complicated and would require a combination of inherent Berry curvature from band structure, chirality-driven AHE due to spin fluctuations, and extrinsic disorders through skew scatterings.

\section{Conclusion}
In conclusion, we report incommensurate magnetism in NdAlGe that onsets at $T_{\text{inc}}~=~6.8(2)~K$ and consists of a modulated SDW with a strong out-of-plane anisotropy and a small helical chiral spin canting of 3(1)\degree. The spin system transitions into a commensurate ferrimagnetic state below $T_{\text{com}}~=~5.1(1)~K$ where the spins form a duu spin structure while keeping the helical spin canting. Similar to NdAlSi, we found that the periodicity of the incommensurate SDW of NdAlGe matches the nesting wave vector between the two of its topologically non-trivial Fermi pockets, which confirms the possibility that Weyl-mediated RKKY interactions could also drive the collective magnetism of NdAlGe. In contrast to NdAlSi, however, NdAlGe has a higher level of disorders, which has a minor effect on its magnetic properties but greatly modifies the transport ones. Effects of disorders in NdAlGe are manifested through the anisotropic ferrimagnetic domains of finite size as well as broad features in the temperature dependence of its magnetic heat capacity, magnetic order parameters, and electrical resistance. In terms of transport, the carrier concentration, mobility, and AHE of NdAlGe are all drastically different from those of NdAlSi. In particular, we characterized an intrinsic as well as an extrinsic AHE regime in NdAlGe that are both absent in the Si analogue. In NdAlGe, we argued that the intrinsic AHE results mainly from its intrinsic Berry curvature, while the extrinsic AHE is tied to disorders and spin fluctuations. The lack of AHE in NdAlSi may be due to differences in the strength of spin-orbit coupling between Ge 4p and Si 3p electrons, and/or an interplay between different mechanisms since its higher RRR may have pushed it towards the clean limit and introduce additional contributions \cite{onoda2006intrinsic,miyasato2007crossover}. Our work thus suggests that Weyl-mediated magnetism is a robust feature of non-centrosymmetric Weyl semimetals NdAl$X$, while the transport properties including AHE in Weyl semimetals can be strongly impacted by extrinsic effects despite the presence of prominent Berry curvature.

\begin{acknowledgements}
H.-Y.Y. thanks Chunli Huang, Hiroaki Ishizuka, Ilya Sochnikov, Christopher Eckberg, Allan MacDonald, Inti Sodemann, Yaroslav Tserkovnyak, and Collin Broholm for fruitful discussions. This material is based upon work supported by the Air Force Office of Scientific Research under Award No. FA2386-21-1-4059. The work at TIFR Mumbai was supported by the Department of Atomic Energy of the government of India under Project No. 12- R$\&$D-TFR-5.10-0100. We acknowledge the support of the National Institute of Standards and Technology, U.S. Department of Commerce. The work at Northeastern University was supported by the US Department of Energy (DOE), Office of Science, Basic Energy Sciences Grant No. DE-SC0022216 and benefited from Northeastern University's Advanced Scientific Computation Center and the Discovery Cluster and the National Energy Research Scientific Computing Center through DOE Grant No. DE-AC02-05CH11231. H.L. acknowledges the support by the National Science and Technology
Council (NSTC) in Taiwan under grant number MOST 111-2112-M-001-057-MY3. The identification of any commercial product or trade name does not imply endorsement or recommendation by the National Institute of Standards and Technology. A portion of this research used resources at the High Flux Isotope Reactor, a DOE Office of Science User Facility operated by the Oak Ridge National Laboratory.
\end{acknowledgements}

\clearpage

\appendix

\section{Crystal structure refinement of NdAlGe}\label{xrd_Refinement}
\begin{figure}[b]
    \includegraphics[width=\columnwidth]{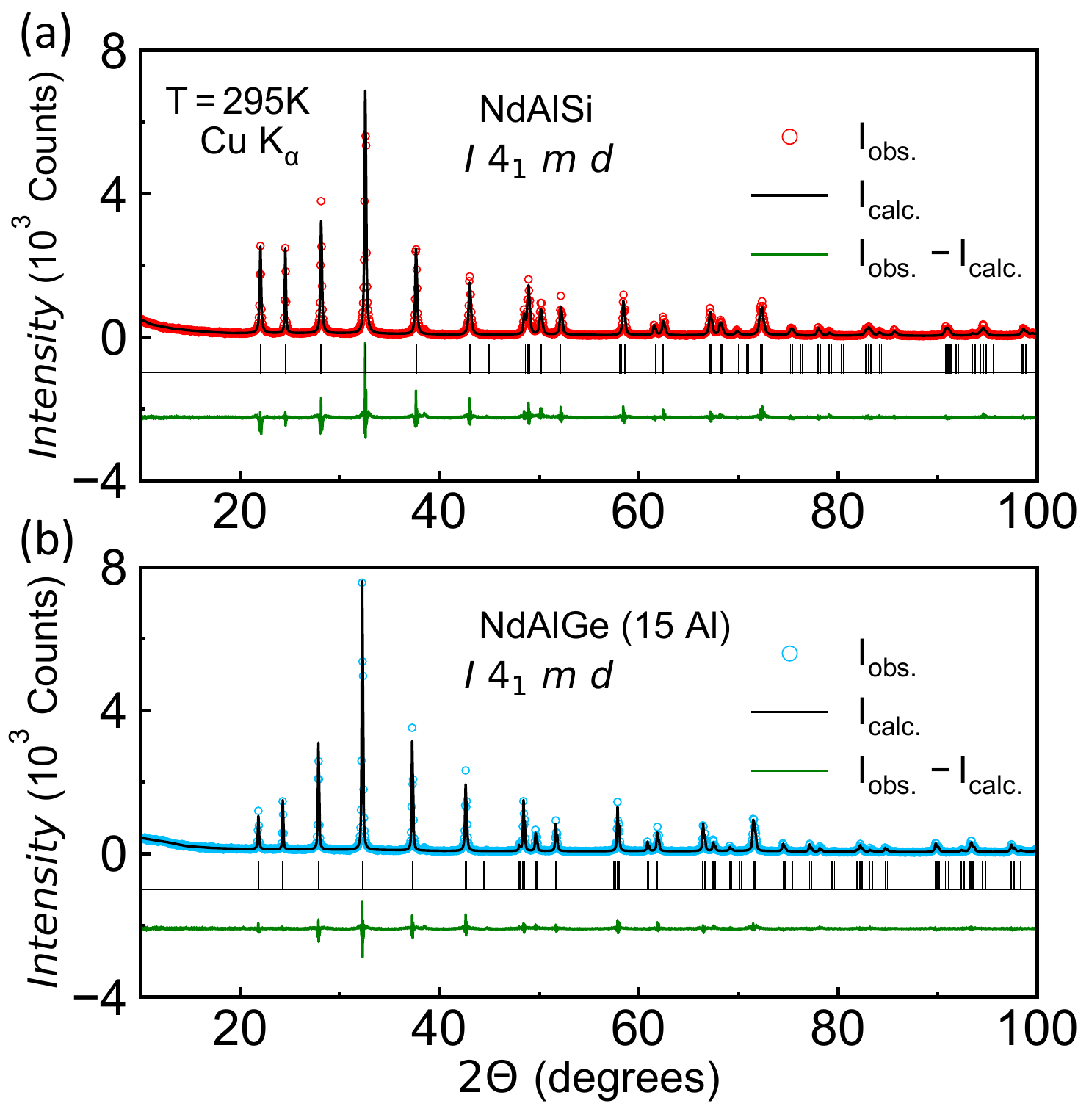}
    \centering
    \caption{Powder XRD refinement of (a) NdAlSi, and (b) NdAlGe samples made with additional Al flux.}
    \label{fig:app_XRD}
\end{figure}

In addition to the powder XRD refinement shown in the main text Fig. \ref{fig:xrd} (NdAlGe made by 10 Al recipe with a refined ratio 1.01:1:0.97), Fig. \ref{fig:app_XRD} shows the refinement of NdAlSi made by 10 Al recipe and NdAlGe made by a 15 Al recipe. The refined ratio for NdAlSi is essentially 1:1:1, and for NdAlGe (15 Al) is $\text{Nd}:\text{Al}:\text{Ge}=0.90:1:0.82$. We note that it is known from neutron scattering that the stoichiometry of NdAlSi single crystals is not exactly 1:1:1 \cite{Gaudet2021}, and we interpret our powder XRD refinement results shown in Fig. \ref{fig:xrd} and Fig. \ref{fig:app_XRD} as follows: relatively speaking, NdAlGe is more non-stoichiometric compared to NdAlSi, and using more Al flux to grow NdAlGe single crystals may result in a higher deficiency in the Nd and Ge sites.

\section{Heat capacity of LaAlGe and NdAlGe}\label{app:app_HC}
\begin{figure}[t]
    \includegraphics[width=0.7\columnwidth]{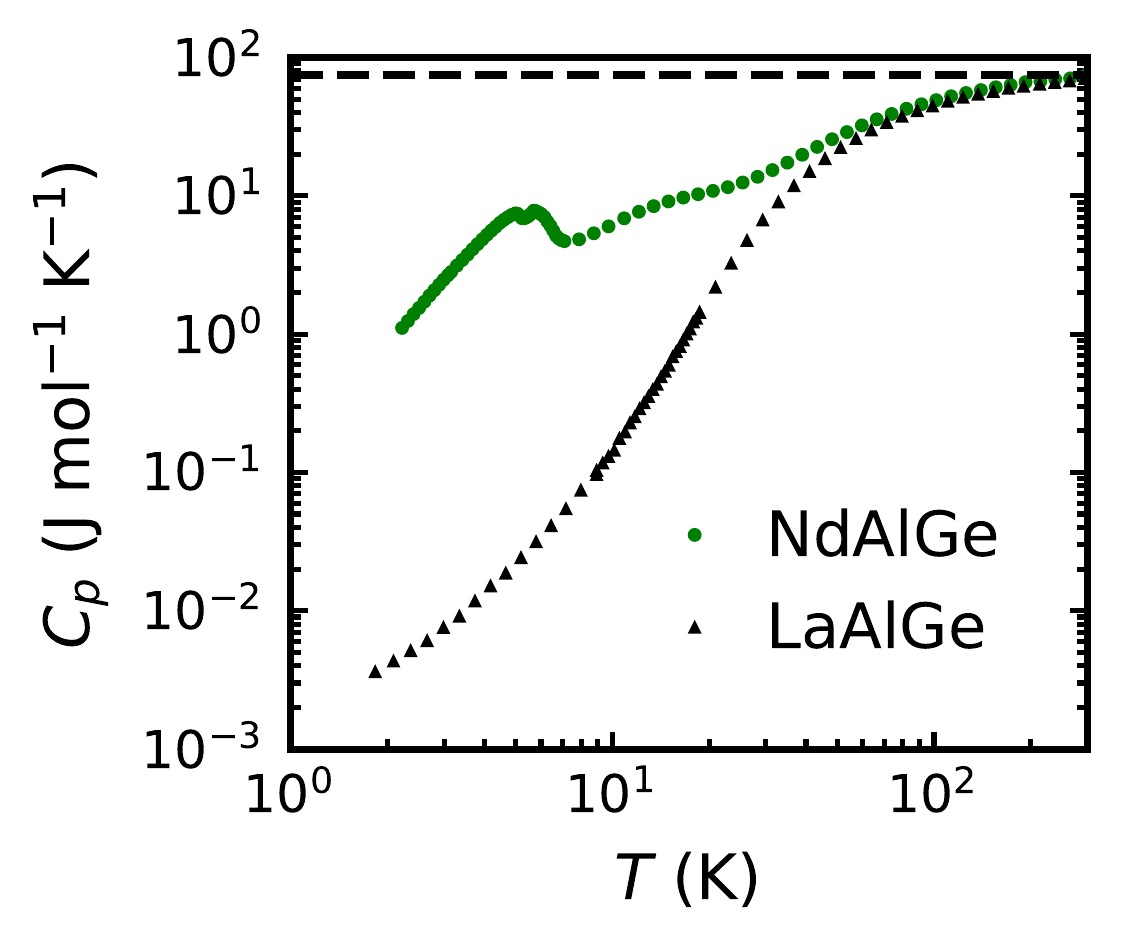}
    \centering
    \caption{Heat capacity data of LaAlGe and NdAlGe. The black dashed line marks the Dulong-Petit limit. The uncertainty in our data is 4$\%$, which comes from the uncertainty of the sample mass.}
    \label{fig:app_HC}
\end{figure}

Fig. \ref{fig:app_HC} shows the heat capacity $C_p(T)$ of both LaAlGe and NdAlGe at zero magnetic field. As expected, the $C_p$ attains the Dulong-Petit limit ($C_p=3 R \times N_\text{ions}$ where $R$ is the gas constant and $N_\text{ions}$ is the number of ions in the material and equals to 3 in NdAlGe and LaAlGe) near the room temperature for both materials. The difference between these two data sets comes from the magnetic specific heat $C_p^{\text{mag}}$, which is analyzed in detail in the main text in Fig. \ref{Fig2_com_inc_Tdep}.

\section{Comparison between the transport properties of NdAlSi and NdAlGe}\label{app:app_rhoxx_rho_yx_comp}
\begin{figure}[b]
    \includegraphics[width=\columnwidth]{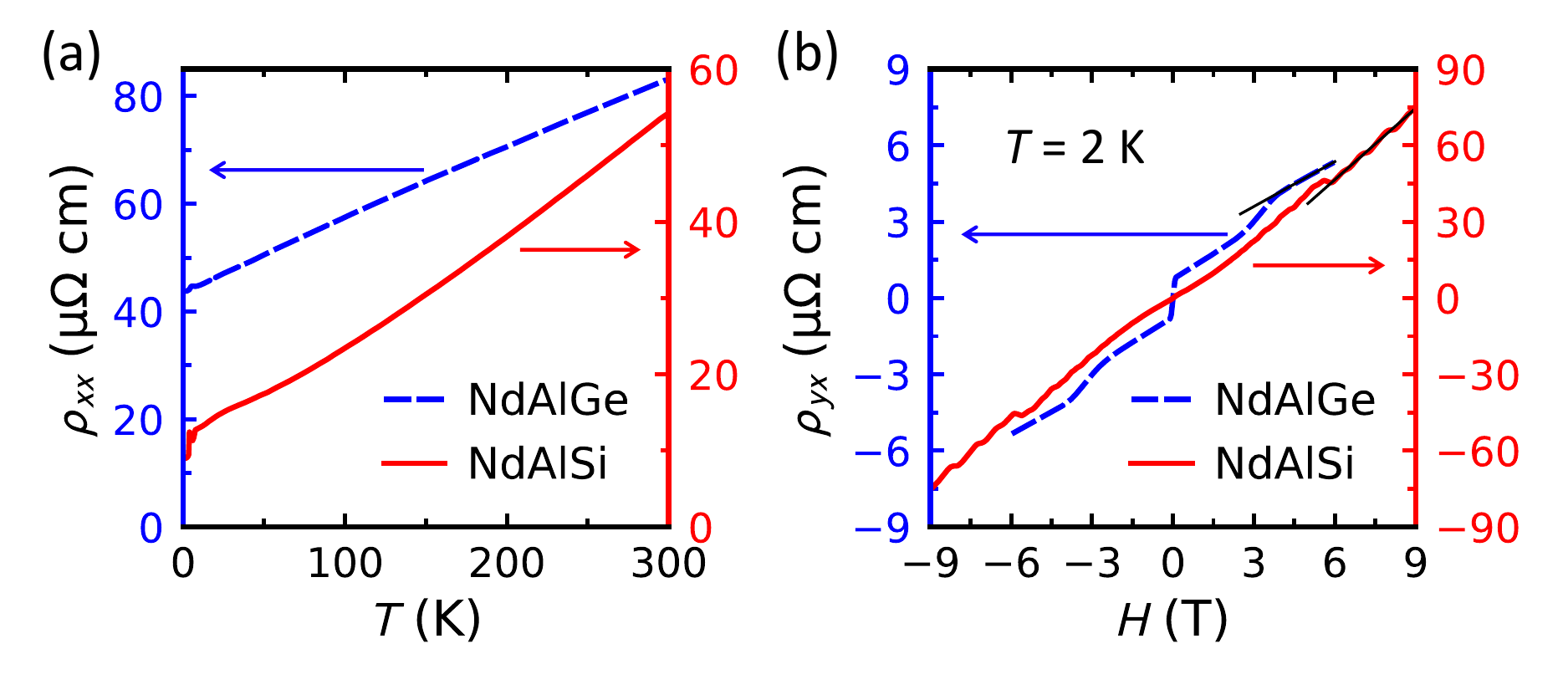}
    \centering
    \caption{(a) $\rho_{xx}(T)$, and (b) $\rho_{yx}(H)$ of NdAlGe (blue dashed lines, left $y$-axis) and NdAlSi (red solid line, right $y$-axis). The high-field part of the data of both materials is fitted to a linear expression (black line in panel (b)) to extract single-band carrier concentrations.}
    \label{fig:app_rhoxx_rho_yx_comp}
\end{figure}

Figure \ref{fig:app_rhoxx_rho_yx_comp} compares $\rho_{xx}(T)$ and $\rho_{yx}(H)$ of NdAlGe and NdAlSi. From Fig. \ref{fig:app_rhoxx_rho_yx_comp}(a), it can be seen that at base temperature $T=2$ K, $\rho_{xx}$ of NdAlGe is much closer to its room-temperature value compared to that of NdAlSi, hence the higher RRR reported in Table \ref{tab:table1}. Figure \ref{fig:app_rhoxx_rho_yx_comp}(b) shows typical $\rho_{yx}(H)$ curves for both materials. For NdAlGe, two plateaus corresponding to the duu and FM magnetic phases can be seen. For NdAlSi, however, although there are also two steps in its magnetization just like NdAlGe \cite{Gaudet2021}, its $\rho_{yx}(H)$ curve is smooth for the first transition and only shows a small discontinuity at the transition field to the FM state. The $\rho_{yx}(H)$ curve is also mildly nonlinear overall such that it is difficult to argue an AHE in NdAlSi (see also Appendix \ref{app:app_rhoyx_MH_NdAlSi}). We obtained the single-band carrier concentrations by fitting a linear line to the high-field part of $\rho_{yx}(H)$ curves for both materials, and used them to calculate their respective single-band mobility from $\rho_{xx}$ at 2~K (see Table \ref{tab:table1}).

\section{$M(H)$ of NdAlGe at different temperatures}\label{app:app_MH}
\begin{figure}[t]
    \includegraphics[width=\columnwidth]{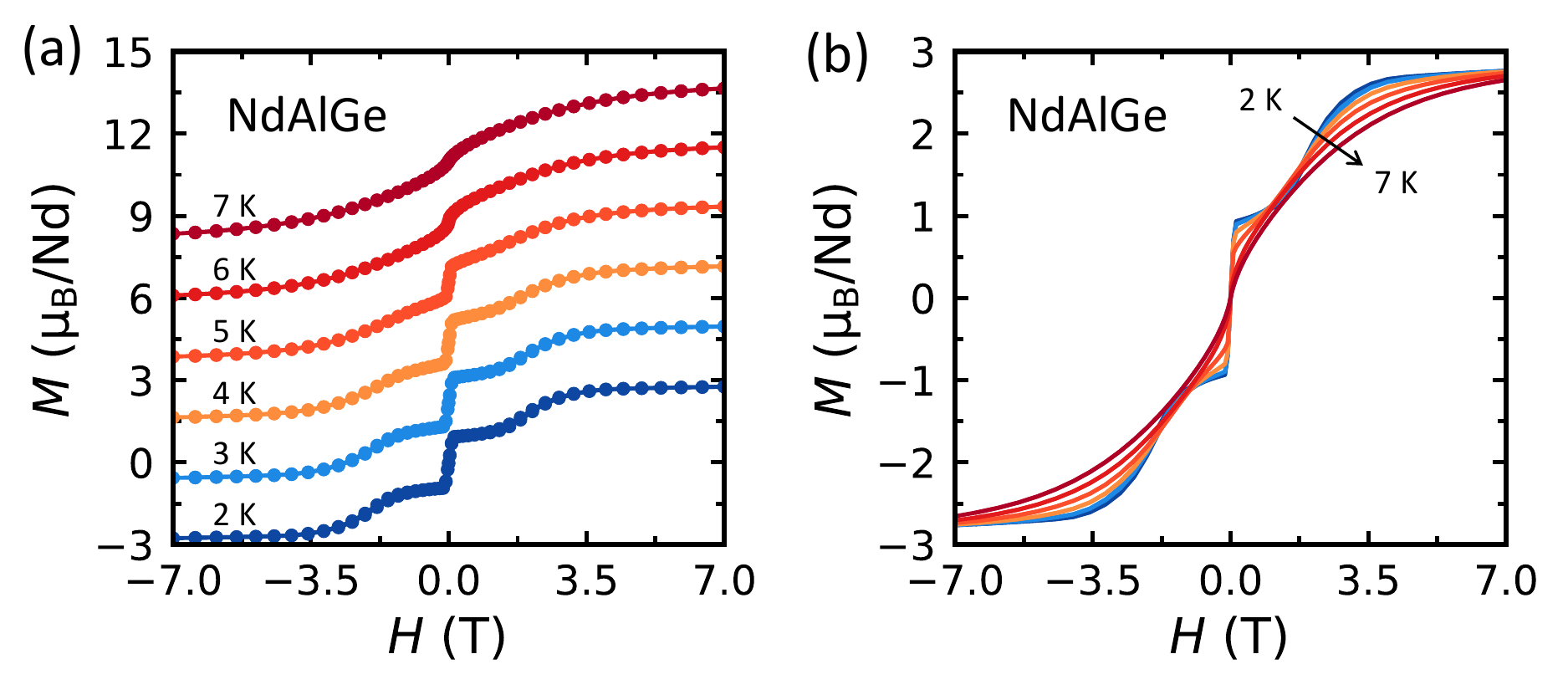}
    \centering
    \caption{(a) $M(H)$ data of NdAlGe recorded at different temperatures. The data at each temperature are vertically shifted away from each other by 2.2 $\mu_\text{B}$/Nd for visibility. (b) The same data as in panel (a), but not shifted.}
    \label{fig:app_MH}
\end{figure}

Figure \ref{fig:app_MH} shows the details of $M(H)$ data of NdAlGe at different temperatures below $T_\text{com}$ and $T_\text{inc}$. The plateaus in the duu and FM states are evident and correspond to the plateaus in $\rho_{yx}(H)$. At high magnetic field, the magnetization converges to the saturated value. The decent saturation of magnetization even at $T=7$ K allows us to extract $\rho^\text{A, FM}_{yx}$ from the high-field plateaus in $\rho_{yx}(H)$ data. We note that the transition field from duu to FM state is not always the same and varies among samples, while the plateaus are persistent before and after the transition.

\section{$\rho_{yx}(H)$ and $M(H)$ of NdAlSi at different temperatures}\label{app:app_rhoyx_MH_NdAlSi}

\begin{figure}[t]
    \includegraphics[width=\columnwidth]{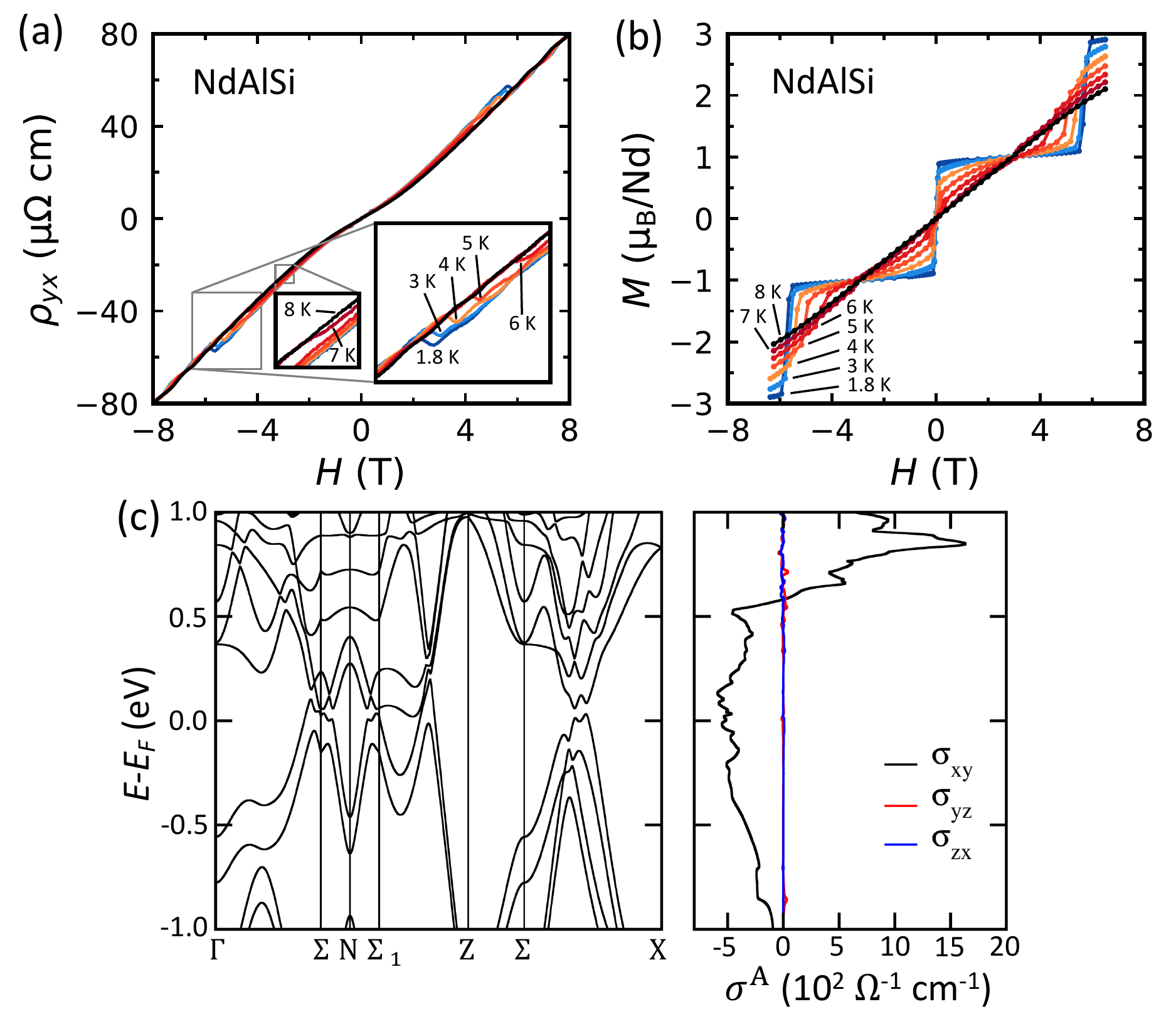}
    \centering
    \caption{(a) $\rho_{yx}(H)$ data of NdAlSi recorded at different temperatures. (b) $M(H)$ data of NdAlSi. (c) Left panel: Band structure of NdAlSi in FM state. Right panel: Anomalous Hall conductivity of NdAlSi, calculated by DFT considering intrinsic Berry curvature. }
    \label{fig:app_rhoyx_MH_NdAlSi}
\end{figure}

Figure \ref{fig:app_rhoyx_MH_NdAlSi}(a) and (b) compare $\rho_{yx}(H)$ and $M(H)$ data of NdAlSi side by side to reveal the absence of Hall resistivity plateaus despite clear plateaus in $M(H)$. Similar to NdAlGe, $M(H)$ of NdAlSi has a low-field transition to duu state and a high-field transition to FM state, each of which results in a plateau in $M(H)$. In the Hall data, however, near $H=0$ T, $\rho_{yx}(H)$ is smooth and featureless; no feature can be associated with the transition to duu state in $\rho_{yx}(H)$. At the transition to FM state, there is a concomitant jump in $\rho_{yx}(H)$ as shown in the inset of Fig. \ref{fig:app_rhoyx_MH_NdAlSi}(a). It is tempting to subtract a smooth background from $\rho_{yx}(H)$ and interpret such a jump as AHE, but this is not feasible here since the $\rho_{yx}$ data in the FM state ($H>7\ \text{T}$) below $T_\text{com}$ actually lie very close to the $\rho_{yx}$ data above $T_\text{com}$ in the same field range. The main reason why they are not exactly on top of each other is that there are quantum oscillations in the low-temperature $\rho_{yx}$ data $(<T_\text{com})$. We note that $M$ significantly drops (more than 30$\%$) as $T$ changes from 1.8 K to 8 K, so if there is a finite AHE, the Hall data recorded at these two temperatures should be quite different from each other due to the proportionality between $M$ and $\rho_{yx}^\text{A}$ \cite{Ye1999}. As a result, we conclude that no Hall plateaus were observed in NdAlSi. The absence of AHE in NdAlSi is interesting because it is in conflict with the AHC calculated by DFT. We performed AHC calculations for NdAlSi in Fig. \ref{fig:app_rhoyx_MH_NdAlSi}(c) by considering the intrinsic Berry curvature, similar to the one shown in Fig. \ref{dft}(f) for NdAlGe. Near $E_F$, there is a persistent AHC ($\sigma_{xy} \sim 500\ \Omega^{-1}\text{cm}^{-1}$), in contrast to the absence of Hall plateaus in Fig. \ref{fig:app_rhoyx_MH_NdAlSi}(a). It would require further theoretical investigation to understand this discrepancy.

\section{Details on the spin structure refinement of NdAlGe}\label{Refinement}
\begin{figure}[t]
    \includegraphics[width=\columnwidth]{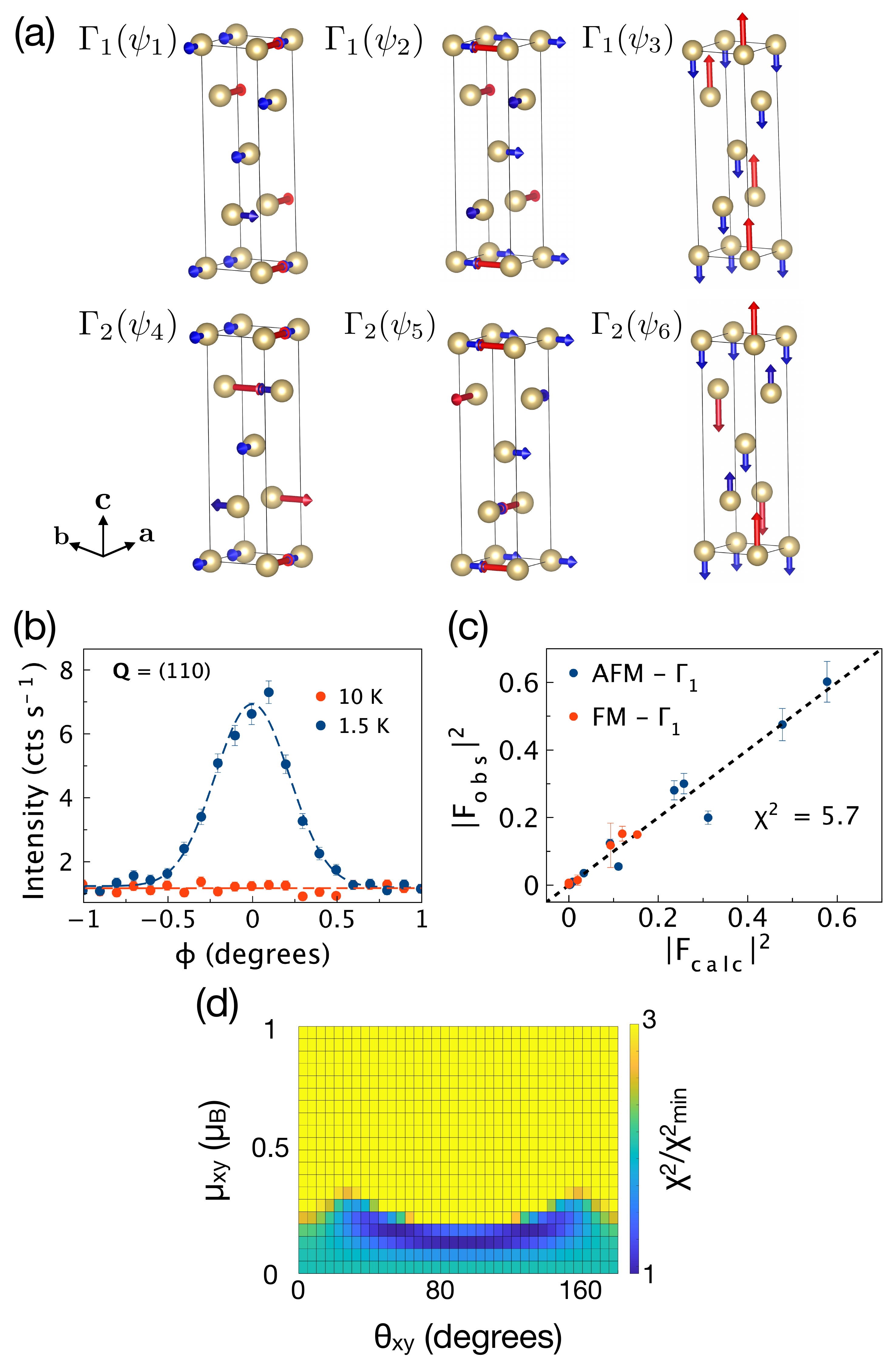}
    \centering
    \caption{(a) shows the real part of the magnetic basis vectors for the AFM spin component of NdAlGe obtained from symmetry analysis. (b) shows rocking scans collected at $\bold{Q}$~=~(1,1,0), which is a higher harmonic originating from the addition of the magnetic $\bold{Q}$=(2/3,2/3,0) and $\bold{Q}$=(1/3,1/3,0) Bragg peaks. The scattered intensity is 100 to 1000 weaker than the magnetic $\bold{Q}$=(2/3,2/3,0) and $\bold{Q}$=(1/3,1/3,0) Bragg peaks. (c) is the final refinement of the spin structure of NdAlGe against both the ferromagnetic (FM) $\bold{k}$=(0,0,0) spin component and the antiferromagnetic (AFM) spin component described by $\bold{k}_\text{AFM1}$ and $\bold{k}_\text{AFM2}$. $|F|^2$ is the neutron structure factor. (d) is the $\chi^2$ value of the neutron diffraction refinement against an in-plane moment $\mu_{xy}$ and its direction within the $ab$ plane $\theta_{xy}$. $\theta$=0 lies along the magnetic ordering vector [1,1,0] direction.}
    \label{DiffApp1}
\end{figure}

We did magnetic refinement of our neutron diffraction data to determine the antiferromagnetic (AFM) spin structure component of the commensurate phase of NdAlGe. To do so, rocking scans at 32 symmetrically nonequivalent Bragg positions were collected at $T=1.5$~K within the manifold of Bragg peaks $\mathbf{Q_+}=\mathbf{G} \pm (\frac{1}{3}+\delta,\frac{1}{3}+\delta,0)$ and $\mathbf{Q_-}=\mathbf{G} \pm (\frac{2}{3}+\delta,\frac{2}{3}+\delta,0)$. Here $\mathbf{G}$ refers to all nuclear allowed Bragg peaks. We performed representational analysis of the \ce{NdAlGe} commensurate magnetism using SARAh refine~\cite{SARAh} and found six possible basis vectors divided into two different irreps ($\Gamma_1$ and $\Gamma_2$)~\cite{rodriguez2001fullprof}, whose real parts are shown in Fig.~\ref{DiffApp1}(a). The two primitive Nd ions located at $\mathbf{r_1}$=(0,0,0) and $\mathbf{r_2}$=(1/2,0,1/4) within the unit cell have spins anti-parallel to each other for spin structures described by $\vec{\psi_1}$+$\vec{\psi_2}$, $\vec{\psi_4}$-$\vec{\psi_5}$, and $\vec{\psi_3}$. These anti-parallel spin structures lead to strong $\mathbf{Q_-}$ peaks and no intensity at $\mathbf{Q_+}$ peaks. On the other hand, spin structures described by $\vec{\psi_1}$-$\vec{\psi_2}$, $\vec{\psi_4}$+$\vec{\psi_5}$, or $\vec{\psi_6}$ have parallel Nd spins at $\mathbf{r_1}$ and $\mathbf{r_2}$. This situation leads to strong $\mathbf{Q_+}$ peaks and no intensity at $\mathbf{Q_-}$ peaks. As seen in Fig.~\ref{BulkDiff}(e) of the main text, we observed intensities at $\mathbf{Q_-}$ positions that are two order of magnitude greater than at $\mathbf{Q_+}$ so the spin structure is dominantly anti-parallel and was refined to an Ising anisotropy ($\vec{\psi_3}$). We also detected weak intensity at $\mathbf{Q_-}$ positions so there's also a weak parallel spin component, which originates from an in-plane spin component $\mu_{xy}$. We did a magnetic refinement against both the $\Gamma_1$ and $\Gamma_2$ manifold and found our refined $\chi_2$ is $\sim$100 times smaller for the $\Gamma_1$ manifold so we concluded that $\Gamma_1$ is the appropriate irrep for NdAlGe. The final refinement is shown in Fig.~\ref{DiffApp1}(c) where the best solution was obtained with $\vec{\psi_1}$=-$\vec{\psi_2}$~=~0.14(2)$\mu_B$ and $\vec{\psi_3}$=3.8(4)$\mu_B$. Furthermore, Fig.~\ref{DiffApp1}(d) shows the $\chi_2$ dependence on both the angle direction of the spin canting within the $ab$ plane ($\theta_{xy}$), and its magnitude ($\mu_{xy}$). In this plot, $\theta_{xy}=0$ is defined to be along the ordering vector direction [1,1,0]. From Fig.~\ref{DiffApp1}(d), we deduced an in-plane spin canting of 0.14(2)$\mu_B/Nd$ oriented 90(20)$\degree$ away from the ordering vector direction, which produces an helical spin canting.  

We determined the spin polarization of the ferromagnetic (FM) $\mathbf{k}=(0,0,0)$ magnetic structure of NdAlSi by collecting rocking scans at 24 symmetrically non-equivalent $\mathbf{k}=(0,0,0)$ Bragg positions covering the $(H,H,L)$ plane. The nuclear and magnetic contributions to the Bragg diffraction were distinguished by collecting rocking scans within both the paramagnetic phase at 10~K and in the commensurate phase at 1.5~K. Symmetry analysis reveals three possible irreducible representations (irreps) to describe the $\mathbf{k}=(0,0,0)$ magnetic structure~\cite{rodriguez2001fullprof}: $\Gamma_1$ and $\Gamma_3$ that respectively correspond to ferromagnetic and antiferromagnetic structures where the spins are oriented along the $c$ axis, and $\Gamma_5$ that describes structures where the spins lie in the $ab$ plane. The Ising ferromagnetic $\Gamma_1$ is the only irrep that matches the symmetry of the antiferromagnetic spin component. Furthermore, $\Gamma_3$ and $\Gamma_5$ respectively produce magnetic Bragg reflections at $\mathbf{Q}=\mathbf{G} \pm (1,1,0)$ and $\mathbf{Q}=(0,0,L)$ positions. As seen in Fig.~\ref{BulkDiff}(e), we did not observe any scattering at $(0,0,L)$ Bragg positions. Also, we only observed extremely weak intensity on the $\mathbf{Q}=\mathbf{G} \pm (1,1,0)$ peaks (such as $\bold{Q}=(1,1,0)$ in Fig.~\ref{DiffApp1}(b)), but these can be understood as arising from the higher harmonics of the AFM order or from the presence of small Nd vacancies. Our final refinement, which is shown in Fig.~\ref{DiffApp1}(c), thus leads to the $\Gamma_1$ structure with $\mu_\text{FM}=0.9(1)\mu_B$.

Related to the phase factor of the spin structure at which neutron diffraction is insensitive, we note that for the Ising component, the spatial variation of the Nd moments is expressed as:
\begin{equation}
\mu_\text{AFM1}(\mathbf{r})=1.9(2)[\exp{(i [(\frac{2}{3}\frac{2}{3}0) \cdot \mathbf{r} +  i \theta])} + c.c.].
\end{equation}
where c.c. stands for complexe conjugate. This expression includes both the $\mathbf{k}=(\frac{2}{3},\frac{2}{3},0)$, and the $\mathbf{k}=(\bar{\frac{2}{3}},\bar{\frac{2}{3}},0)$ components as required for the magnetic moment to be real for all $\mathbf{r}$. While the diffraction pattern is independent of $\theta$, the real space spin structure does depend on $\theta$. For $\theta~=\pi$, the spin structure can be described as ($0$-up-down) where $0$ means there is no net magnetization on this site, whereas a $\theta~=0$ phase shift leads to an (up-down-down) spin structure. Within the commensurate phase, once the FM component of the structure is added ($\mu_\text{FM}$~=~0.9(1)$\mu_B$), $\theta~=0(6)\degree$ is the only phase that allows for all the Nd moments to not exceed the 2.80(5)$\mu_B$ saturated moment determined by the out-of-plane magnetization data.

\section{SANS data with 12 \angstrom~neutrons}\label{SANSApp}
\begin{figure}[t]
    \includegraphics[width=\columnwidth]{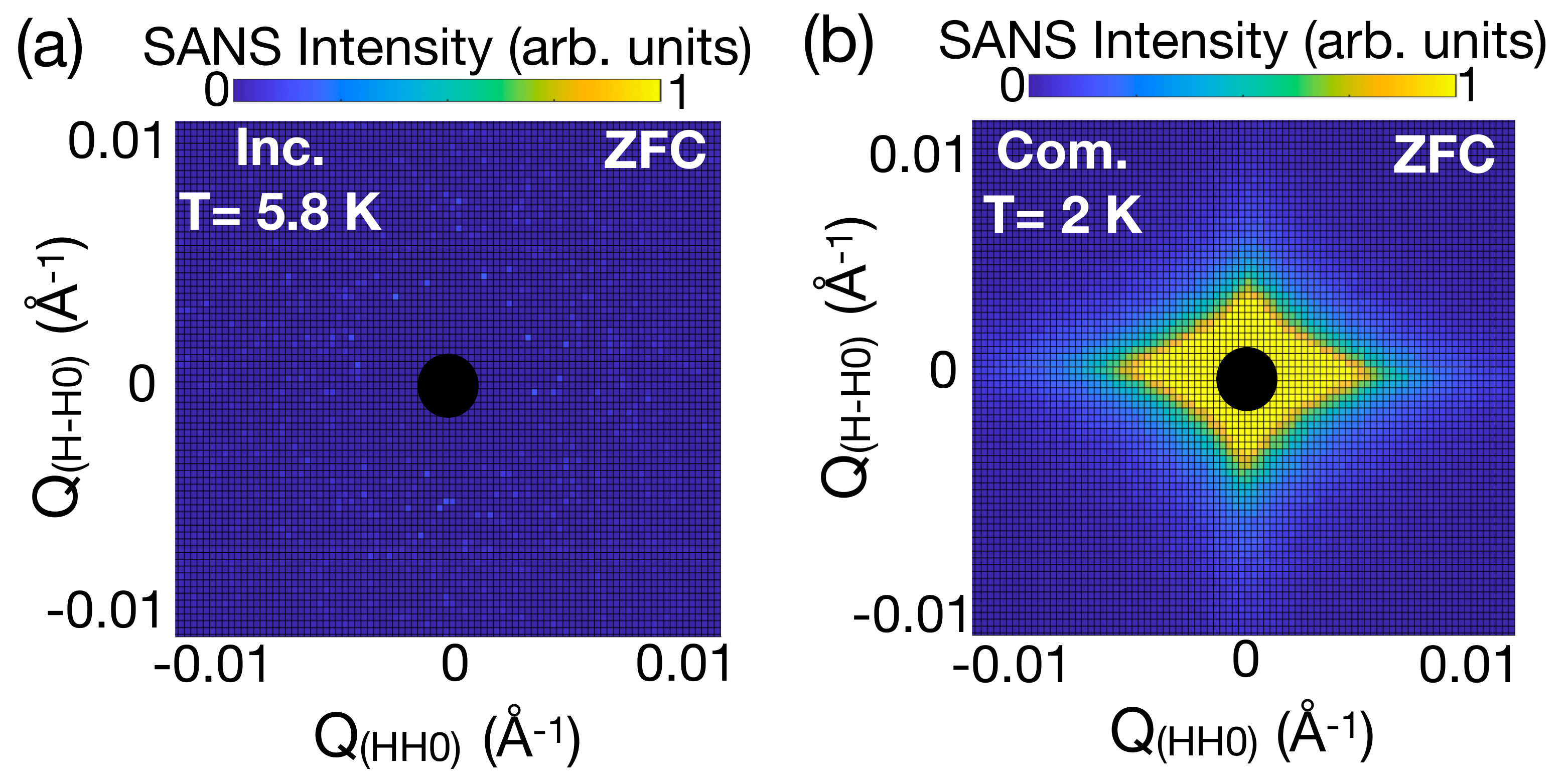}
    \centering
    \caption{(a) and (b) correspond to the zero-field cool (ZFC) $12~\angstrom$ SANS data respectively collected within the magnetic incommensurate phase (5.4~K) and the commensurate phase (2~K) of NdAlGe.}
    \label{SANS2}
\end{figure}
As discussed in the main text, we collected SANS pattern with 12 \angstrom~incident neutrons for various temperatures ranging from 10 to 2~K. Representative SANS pattern collected within both the magnetic incommensurate and the magnetic commensurate phase of NdAlGe are respectively shown in Fig.~\ref{SANS2}(a) and (b). As seen from the absence of SANS scattering within the incommensurate phase, the 12~\angstrom~data set was used to isolate the SANS scattering from the $\bold{Q}=0$ cross pattern by summing over all the scattering detected in such data set. The temperature dependence of the cross pattern extracted this way is presented in Fig.~\ref{BulkDiff}(b) of the main text (blue dots).

\bibliography{projects-RAlX-NdAlGe_project}
\bibliographystyle{apsrev4-1}

\noindent

\end{document}